 \documentclass[aps,reprint,superscriptaddress]{revtex4-1}

\usepackage{graphicx} 
\usepackage{amsmath,bm}	
\usepackage{amsmath,amssymb}
\usepackage{graphicx,epsfig,sidecap,wasysym}
\usepackage{color,epstopdf}
\usepackage[colorlinks,bookmarks=false,citecolor=blue,linkcolor=blue,urlcolor=blue]{hyperref}
\newcommand{\Z}{\mathbb{Z}}
\usepackage{xcolor}
\begin{document}
\title{Effects of a single impurity in a Luttinger liquid with spin-orbit coupling   } 

 \affiliation{  Physics Department, National Research University Higher School of Economics, Moscow, 101000, Russia} 
 \affiliation{L. D. Landau Institute for Theoretical Physics, Chernogolovka, Moscow region 142432, Russia }
 \affiliation{Russian Quantum Center, Skolkovo, Moscow 143025, Russia}
\author{M. S. Bahovadinov}
\affiliation{  Physics Department, National Research University Higher School of Economics, Moscow, 101000, Russia} 
\affiliation{Russian Quantum Center, Skolkovo, Moscow 143025, Russia}
 \author{S. I. Matveenko}
 \affiliation{L. D. Landau Institute for Theoretical Physics, Chernogolovka, Moscow region 142432, Russia }
\date{\today}
\affiliation{Russian Quantum Center, Skolkovo, Moscow 143025, Russia}

\begin{abstract}
In quasi-1D conducting nanowires spin-orbit coupling destructs spin-charge separation, intrinsic to Tomonaga-Luttinger liquid (TLL).   We study renormalization of a single scattering impurity in a such liquid.  Performing bosonization of low-energy excitations and exploiting perturbative renormalization analysis we extend the phase portrait in $K_\sigma - K_\rho$ space, obtained previously for TLL with decoupled spin-charge channels. 
   
\end{abstract}

\maketitle



\section{Introduction}  
Low-energy  excitations of the one-dimensional conductors have collective phonon-like character, formalized within the Tomonaga-Luttinger liquid (TLL) theory ~\cite{TomonagaTLL,LuttingerTLL,GogolinTLL,   HaldaneTLL}. The TLL formalism  correctly predicts algebraic decay of single-particle correlations with exponents depending on the strength of the (short-range) electron-electron (e-e) interactions. Due to the collective nature of these excitations the interparticle interactions drastically change the low-energy physics leading to fractionalization of constituent carriers~\cite{Glazman,KLeHurFrac,KamataFrac}, vanishing density of states and power law singularity of the  zero-temperature momentum distribution at the Fermi level~\cite{ VoitTLL,GiamarchiTLL,JeckelmanDOS}.  Another surprising prediction of the TLL theory is related to the effects of a single scattering impurity on transport properties at low temperatures, presented in prominent series of works in Refs.~\cite{KaneImpurity,NagaosaImpurity,KaneImpurity2}. Perturbative renormalization group (RG) studies have shown that if the electron-electron interactions of the single-channel TLL is attractive (with the TLL parameter $K>1$), impurity is irrelevant in the RG sense due to the pinned superconducting fluctuations. On the other hand, if the e-e interactions have a repulsive character ($K<1$), the backscattering effects are relevant and grow upon the RG integration of short-distant degrees of freedom. At $T=0$ the system is effectively decoupled into two disjoint TLLs for arbitrarily weak impurity potentials. The latter implies a metal-insulator transition with vanishing two-terminal dc-conductance $G=0$ at $T=0$, whereas at finite temperatures it exhibits power-law dependence on T, i.e  $G \sim  T^{2/K-2}$. 
These results are in sharp constrast with the result in the non-interacting limit ($K=1$), where the impurity is marginal in the RG sense (in all orders of RG) and the transmission coefficient ${\cal T}$ depends on the impurity strength with Landauer conductance $G={\cal T} \frac{e^2}{h}$. Experimental confirmation of these effects was performed recently, using remarkable quantum simulator based on a hybrid circuit~\cite{QuantumSimulator}.    
 \begin{figure}[t]
\includegraphics[width=8.5cm ]{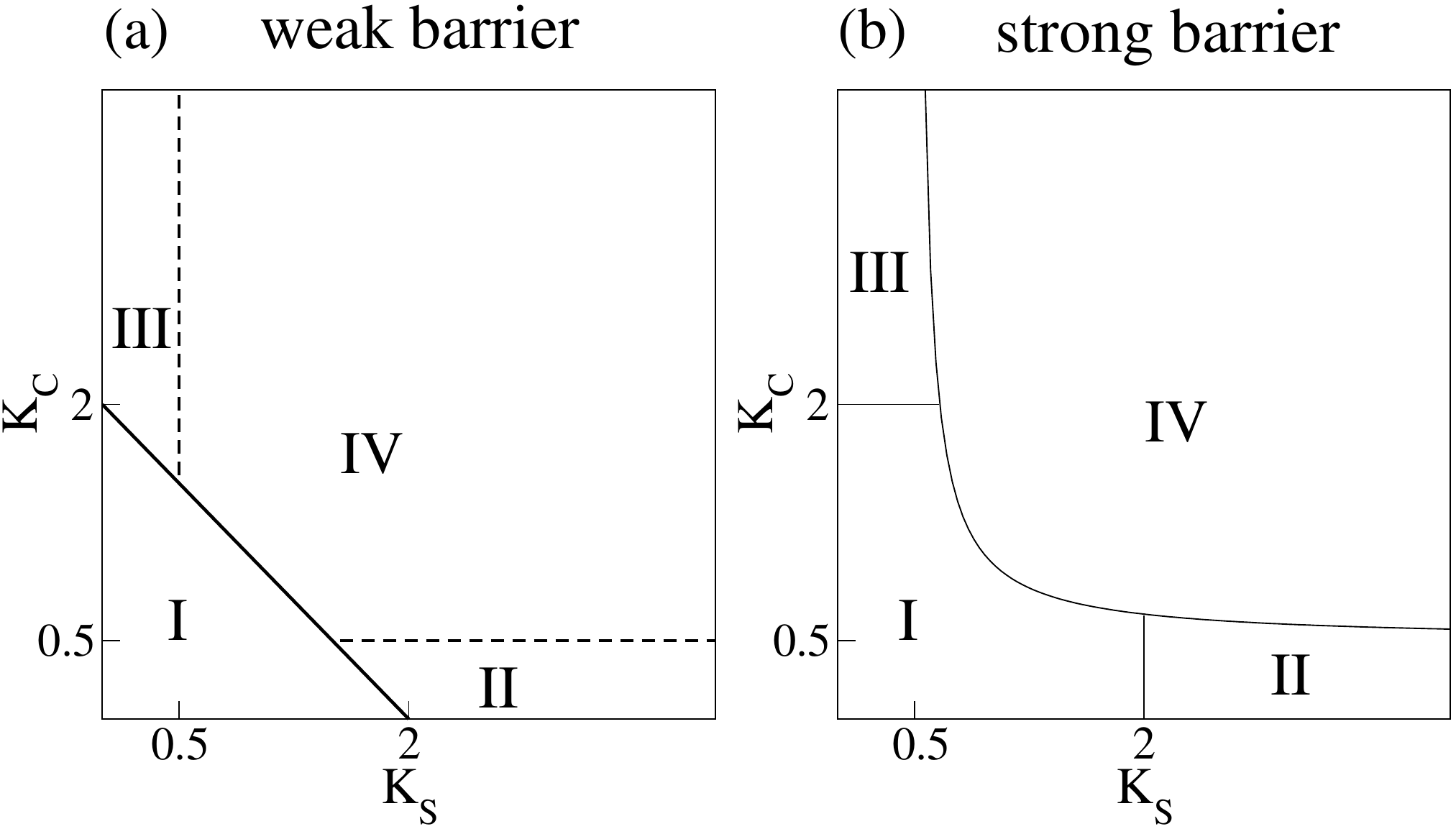}%
\caption{\label{fig:fig1} Phase portrait at $T=0$ for TLL with separated spin and charge channels.  Results are obtained from perturbative RG studies of (a) a weak scattering barrier or (b) a weak link. In region I the backscattering of carriers from a weak barrier (tunneling through a link) is relevant (irrelavant) and flows to the strong-coupling (to the weak-coupling) fixed point in both channels. The opposite occurs in the regions IV. In the remained regions the mixed phase is realized, where one of the channels is insulating and the other is conducting.
} 
\end{figure}
  \label{sec:phasePortrait}

In TLLs composed of spin-1/2 electrons~\cite{RealisticTLLChiralQHE,RealisticTLLChiralQHE2,RealisticTLLNanotube,RealisticTLLWire} spin and charge degrees of freedom are decoupled~\cite{JompolSCS,BocquillonSCS, HashisakaSCS,Vekua,Brazovskii}, in contrast to higher dimensional counterparts. Each channel has individual bosonic excitations with the proper velocity and carry the corresponding quantum number separately.  
	 However, this decoupling of modes is not characteristic for all TLLs. Particularly, the earlier studies have shown that spin-charge coupling (SCC) occurs in the TLL subjected to a strong Zeeman field ~\cite{Aoki1996} or with strong spin-orbit interactions \cite{MorozTheory,MorozExpr,Iucci}. The latter is typically expected in nanowires, where electrons in transverse directions are confined, whereas in the other direction they move freely. Spin-orbit coupling (SOC) in these systems plays an important role on the realization of spintronic devices~\cite{Datta1990,Frolov2010}. Thus, an interplay of (large) SOC and (large) e-e interactions in the TLL regime is an interesting question and has gained both theoretical and experimental attention in recent years~\cite{FasthSOC,SatoExperiment, SatoTheory}.
	
Impurity effects in conventional TLLs with decoupled channels were studied in the original works \cite{ NagaosaImpurity,KaneImpurity2}, and can be summarized by phase portrait at $T=0$, presented in Fig.~\ref{fig:fig1}.  RG studies imply that electrons scattering on a weak potential barrier [Fig.~\ref{fig:fig1}(a)] are fully reflected when $K_c+K_s<2$ (region I). In region II (III) charge (spin) quanta are fully reflected, while the spin (charge) ones are transmitted. The backscattering terms scale to the weak-coupling fixed point in region IV, so electrons fully transmit through the barrier. The same physical picture is obtained from the perturbative analysis of the tunneling events between disconnected wires [Fig.~\ref{fig:fig1}(b)]. In region I  $(\frac{1}{K_c}+\frac{1}{K_s}<2,  K_c<2,  K_s<2)$ all hopping events are irrelevant, so the system renormalizes into two disconnected wires. In region II (region III) only  tunneling of spin (charge) is relevant. Numerical confirmation of these results was presented using path-integral Monte-Carlo methods ~\cite{NumericalStudyPhasePortrait}. 
 
The modification of the phase portrait in the presence of SCC has not yet been considered, although there are several studies which partially addressed this question in different aspects~\cite{Matveev, KimuraResonantZeeman}. Particularly, impurity effects on transport properties in the case of the SCC caused by SOC were recently studied~\cite{SatoExperiment,SatoTheory}.  As our main motivation of this work we consider modification of the phase portrait in the presence of SOC. We also show that the  spin-filtering effect conjectured for this model in Ref.~\cite{KimuraResonantZeeman}, is not exhibited.     

The paper is organized as follows: In Section II we present the model of our study and set the necessary formalism and conventions. We next present a generalized approach to tackle the impurity problem for a finite SOC in Section III. In Section IV we present our main results and discussions.  Concluding remarks are given in Section V.
\section{Model and methods}   
In this section we set up our used conventions and terminology and present the model of our study.
 We approach the problem using the standard Abelian bosonization technique with the consequent perturbative RG analysis of impurity terms.

\begin{figure}[t]
\includegraphics[width=8.5cm ]{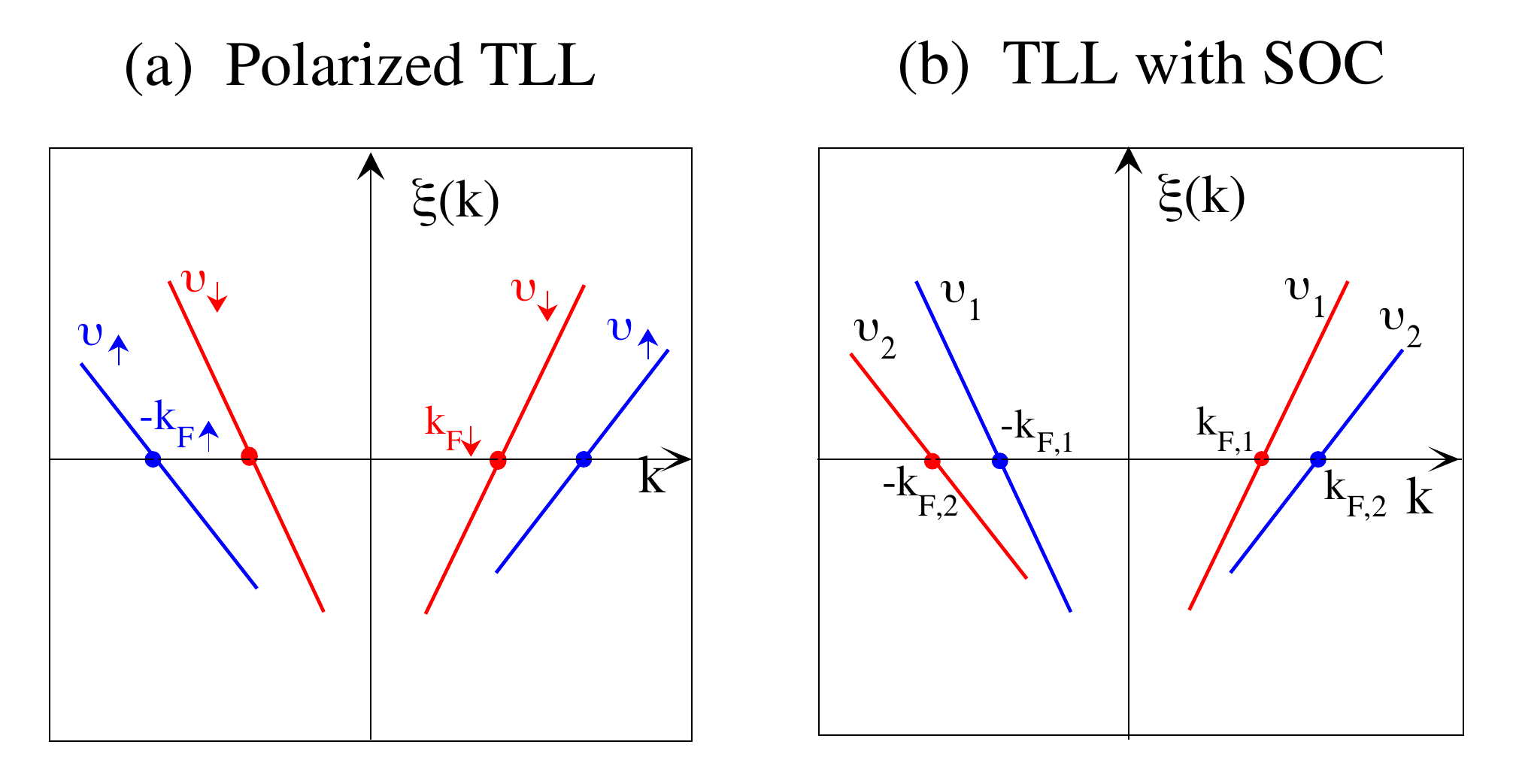}%
\caption{\label{fig:fig2} Linearized electronic excitation spectrum corresponding to the (a) TLL in a strong Zeeman field and (b) TLL with SOC. In both models spin-charge separation is violated by the mixing terms given in Eq.~(\ref{Hmixing}) and Eq.~(\ref{HmixingB}).
} 
\end{figure}
  \label{sec:phasePortrait}
Bosonization of low-lying modes relies on the assumption of linear electronic spectrum with the corresponding right (R) and left (L) movers for carriers with both spin projections  (see Fig. 2). Fermionic field operators for each branch and spin component can be expressed via the bosonic displacement and phase fields:
\begin{equation}
\psi_{\sigma,\mu}=\frac{\hat{F_\sigma}}{\sqrt{2\pi a}}e^{i\mu k_{F,\sigma}x}e^{i\mu \sqrt{\pi}(\phi_\sigma(x) - \mu \theta_\sigma(x))} 
\end{equation}
with $\sigma \in \left\lbrace \uparrow \equiv +1 , \downarrow \equiv -1 \right\rbrace $ and $\mu \in \left\lbrace  R  \equiv +1  ,L \equiv -1 \right\rbrace  $. 
 The dual fields satisfy the commutation relations: 
\begin{equation}
[\phi_\sigma(x), \Pi(x^\prime)_{ \sigma^\prime} ] =i \delta_{\sigma,\sigma^\prime} \delta(x-x^\prime),  
\label{CommRel}
\end{equation}
where $\Pi_\sigma(x) = \nabla \theta_\sigma(x)$ is a conjugate to $\phi_\sigma(x)$ momentum.  
 The UV cutoff $ a $ of the theory is on the order of inter-atomic distances. As in the standard literature, the Klein factors $\hat{F}_\sigma$  guarantee anti-commutation of fermionic fields with different spin orientations.
    
The second exponent can also be expressed in terms of constituent bosonic ladder operators in momentum space ($l$ is the length of the system)  
\begin{equation}
 i\mu \sqrt{\pi}(\phi_\sigma(x) - \mu \theta_\sigma(x))  = \sum_{\mu q>0} A_q\left(b_{q,\sigma}e^{iqx}-b^\dagger_{q,\sigma}e^{-iqx}  \right)
\end{equation}
with $A_q=\sqrt{\frac{2\pi}{l|q|}}e^{-a|q|/2}$ and $[b_q,b^\dagger_{q^\prime}]=\delta_{q,q^\prime}$, $q=\frac{2\pi}{l}$. 
  
We hereafter use the spin and the charge basis as the canonical basis, i.e we work with the fields $\phi_{ c,s}=\frac{\phi_\uparrow \pm \phi_\downarrow}{\sqrt{2}}$  and   $\Pi_{ c,s}=\frac{\Pi_\uparrow \pm \Pi_\downarrow}{\sqrt{2}}$, which also satisfy Eq.~(\ref{CommRel}). Hereafter, the subbscripts "s" and "c" stand for spin and charge degrees of freedom.
The fluctuations of the spin and the charge density are given by,
\begin{equation}
\rho_{s }(x)=\sqrt{\frac{2}{ \pi  }}\sum_\sigma \sigma\partial_x \phi_\sigma 
\end{equation}
and
\begin{equation}
\rho_{c }(x)=\sqrt{\frac{2}{ \pi  }}\sum_\sigma  \partial_x \phi_\sigma. 
\end{equation}
 To get similar expressions for the current, one uses the transformation $\partial_x\phi_\sigma \rightarrow \partial_x \theta_\sigma $. In the following, we imply normal ordering with respect to Dirac sea, whenever it is needed and avoid $:():$ symbol.

 \subsection*{Model}
  The important effect of SOC in quantum nanowires is band distortion, which is usually considered within the two-band model~\cite{Governale}. This distortion causes the velocity difference $\Delta=v_1-v_2$, pronounced in Fig.~\ref{fig:fig2}(b) in the approximation of linearized spectrum. Initially proposed in Refs.\cite{MorozExpr, MorozTheory}, the model can be realized by tuning chemical potential, filling only the lowest subband. It should be noted that the spin orientation of carriers moving in the same direction can be tuned also from the parallel~\cite{Zulicke} to anti-parallel ~\cite{MorozTheory} by band-filling~\cite{SatoTheory}. In this work we consider anti-parallel spin orientation, as shown in~Fig.~\ref{fig:fig2}(b). 
 
 The non-interacting Hamiltonian of the model with linearized excitation spectrum is
  \begin{equation}
  \begin{aligned}
 H_0 & =-iv_1\int dx \left( \psi^\dagger_{R,\downarrow}(x) \partial_x\psi_{R,\downarrow} (x) -  \psi^\dagger_{L,\uparrow}(x) \partial_x\psi_{L,\uparrow} (x)\right)  \\        &  -iv_2\int dx \left( \psi^\dagger_{R,\uparrow}(x) \partial_x\psi_{R,\uparrow} (x) -  \psi^\dagger_{L,\downarrow}(x) \partial_x\psi_{L,\downarrow} (x)\right),  
 \end{aligned}
 \end{equation}
  with distinct Fermi velocities $v_{1}$ and $v_{2}$. As it is clear from the electronic spectrum, the model has broken chiral symmetry, but the time-reversal symmetry is preserved.
  
  Corresponding excitations can be rewritten equivalently in the bosonic language $b_q$ (neglecting zero-energy modes) 
 \begin{equation}
 \begin{aligned}
 H_0 &=v_1 \left( \sum_{q>0}|q|b^\dagger_{q,\downarrow} b_{q,\downarrow} + \sum_{q<0}|q|b^\dagger_{q,\uparrow} b_{q,\uparrow} \right)\\
     &+ v_2 \left( \sum_{q>0}|q|b^\dagger_{q,\uparrow} b_{q,\uparrow} + \sum_{q<0}|q|b^\dagger_{q,\downarrow} b_{q,\downarrow} \right).
 \end{aligned}
 \end{equation}
Collecting the common terms, we obtain  
   \begin{equation}
   H_0 =\frac{v_F}{2}  \sum_{q \neq 0, \sigma}|q|b^\dagger_{q,\sigma} b_{q,\sigma} + \frac{\Delta}{2} \sum_{q\neq 0, \sigma}q \sigma b^\dagger_{q,\sigma} b_{q,\sigma} 
   \label{HamBq}
      \end{equation} 
with $\Delta= v_1-v_2$ and $v_F= v_1+v_2 $. The second term vanishes for vanishing SOC and has the following form in the spin-charge basis:
\begin{equation}
 H_{mix}=-\frac{\Delta}{2} \sum_{q\neq 0}     q \left[ b^\dagger_{q,s}b_{q,c}+b^\dagger_{q,c}b_{q,s} \right]. 
\end{equation}
This term with finite $\Delta \neq 0 $ violates the spin-charge separation and demands more general approach. 

We consider the interacting theory within the generalized g-ology approach and do not impose any constraint on TLL parameters $K_\nu$, $ \nu \in \left( s,c \right)$. Within this generalization, the coordinate space representation of the interacting Hamiltonian in the spin-charge basis has quadratic Gaussian form: 
 \begin{equation}
H_{SOC} = \sum_{\nu=s,c} \frac{v_\nu }{2}\int  \left[ \frac{1}{K_\nu}\left( \partial_x\phi_\nu\right)^2+K_\nu(\partial_x\theta_\nu)^2 \right] dx +H_{mix}.
\end{equation}
 
The SCC term is expressed as follows:   
\begin{equation}
H_{mix}=\frac{\Delta }{2} \int \left[\partial_x \phi_s \partial_x \theta_c + \partial_x \phi_c\partial_x \theta_s \right] dx.
\label{Hmixing}
\end{equation}
 For vanishing SOC, the chiral symmetry is restored, with full $SU(2)$ symmetry and $K_s=1$.
 
 In principle, one has to include also the backscattering term to the Hamiltonian,
 \begin{equation}
 H_{BS}=\frac{g_s}{2(\pi a)^2} \int dx \cos(\sqrt{8\pi}\phi_s).
 \end{equation}
 In the parameter space of its relevancy, this term opens a gap in the spin channel via the Berezinskii-Kosterlitz-Thouless mechanism. There are several works which have addressed the relevancy of this term for repulsive e-e interactions~\cite{MorozTheory, Gritsev,Schulz,Kainaris}. However, in a recent experiment on InAs nanowires with a \textit{strong} SOC, no sign of spin gap is observed~\cite{SatoExperiment}. Based on the phenomenology, we neglect all such terms within the whole parameter space. We also emphasize that impurity effects in the system with gapped spin channel were also previously studied \cite{Kainaris,KainarisImp}.
 
For perturbative analysis of impurity effects, the imaginary-time Euclidean actions for the displacement fields  can be obtained by integrating out the quadratic phase fields. In the $\vec{ x}=( x , \tau)$ space  it takes the following form:
\begin{equation}
S^\phi_\nu=\frac{1}{2v} \int dx d\tau  \left((\partial_\tau \phi_\nu)^2 +\tilde{v}_\nu^2  (\partial_x \phi_\nu)^2\right)  
\label{actionPhiFull}
\end{equation} 
with $v=v_sK_s=v_cK_c$, guaranteed by Galilean invariance of the model with $\Delta=0$.  The renormalized velocities are:
 \begin{equation}
\tilde{v}_\nu^2=v_\nu^2 d_\nu,
\label{veltildeDef}
\end{equation}
\begin{equation}
d_\nu=\left(1-\frac{\Delta^2}{4v_\nu^2} \right).
\end{equation}
The contribution from the mixing term is,  
\begin{equation}
S^\phi_{mix}=\frac{ i \Delta  }{2v} \int  dx d\tau  \left( \partial_x \phi_s\partial_\tau \phi_c + \partial_x \phi_c\partial_\tau \phi_s\right).
\label{actionPhimix}
\end{equation}
 
 Euclidean actions for the phase field fluctuations can be obtained in a similar fashion, 

\begin{equation}
S^\theta_\nu=\frac{v}{2} \int dx d\tau  \left( \frac{1}{v_\nu^2} (\partial_\tau \theta_\nu)^2+d_{-\nu}(\partial_x\theta_\nu )^2  \right)  
\end{equation}
along with the mixing $\theta$-term 
\begin{equation}
S^\theta_{mix}=  \frac{i v\Delta}{2} \int  dx d\tau   \left(\frac{1}{v_c^2} (\partial_x\theta_s\partial_\tau\theta_c)+\frac{1}{v_s^2}(\partial_x\theta_c\partial_\tau\theta_s) \right).
\end{equation}
 
 For $\Delta=0$, one correctly obtains the standard expressions for actions $S^{\phi}_{\nu}$ and $S^{\theta}_{\nu}$.

\subsection*{Impurity bosonization }
We consider a single impurity embedded at the origin $x=0$. This impurity causes backscattering of carriers  
\begin{equation}
H_{bs} = V_{bs} \sum_\sigma \left( \psi^\dagger_{R,\sigma} \psi_{L,\sigma} +h.c. \right),
\label{ImpHam}
\end{equation}
where $V_{bs}$ is the Fourier component of the backscattering potential $ V(k_{F,1}+k_{F,2})$. The forward-scattering term can be always gauged out~\cite{GogolinTLL}.
 
The bosonized expression of the backscattering term leads to the boundary sine-Gordon model in the spin-charge basis: 
\begin{equation}
H_{bs}=\frac{2V_{bs}}{\pi a}\cos(\beta_I \phi_s(x=0))\cos(\beta_I \phi_c(x=0)) 
\end{equation}
with 
$\beta_I^2=2\pi$. This term also couples spin and charge degrees of freedom at the impurity point.  We note that in the bosonization process of this term we do not take into account Klein factors $\hat{F}_\sigma$, since their effect is irrelevant within our model with all terms included terms in this work. The corresponding backscattering action takes the following form:
\begin{equation}
S_{b }=\frac{2V_{bs}}{\pi a} \int_0^\beta d\tau \cos(\beta_I \phi_s(0,\tau))\cos(\beta_I \phi_c(0,\tau)) 
\label{ImpAct1}
\end{equation}
 $\beta=\frac{1}{ T}$ is the inverse temperature and should not be confused with $\beta_I$. We hereafter assume $V_{bs}/\Lambda  \ll 1$, where $\Lambda $ in the effective bandwidth for both channels.  

For a complete analysis, pertubative study of the backscatteting action Eq.~(\ref{ImpAct1}) should be accompanied with the RG study in the strong barrier limit. For this purpose, we consider two disjointed wires and treat interwire tunneling term perturbatively.  
 The most relevant tunneling term has the following contribution to the total action,
 \begin{equation}
S_t=\frac{t}{\pi a} \int_0^\beta d\tau \cos(\beta_I \theta_s(0,\tau))\cos(\beta_I \theta_c(0,\tau)),
\label{TunnelingImp}
\end{equation}
where $t$ is the bare tunneling amplitude through the weak link. 
 \section{Effective baths actions}
At this point it is worth noting that the previous bosonization studies of TLL with SCC~\cite{Aoki1996,KimuraResonantZeeman,SatoTheory} were performed using basis rotation approach, where one transforms the initial basis to the new spin-charge basis with decoupled channels and new renormalized velocities and $K_\nu$ parameters. This approach usually results on redundant expressions for model parameters and complicates further analysis. Instead, we follow a generic approach \cite{GogolinTLL}  without performing a basis rotation. As we show below, this procedure is not required.  We next trace out all space degrees of freedom except for the impurity point. The low-lying excitations of the fields away from the impurity act as a dissipative bath reducing the problem to an effective  0D field theory of a single Brownian quantum particle in a 2D harmonic impurity potential~\cite{Zwerger1985}. 
 
We follow the standard procedure~\cite{GogolinTLL,NagaosaImpurity} to get effective bath actions for the displacement fluctuation. The same procedure applies for the phase fields.  

 The integration of the bulk degrees of freedom is done using the standard trick of introducing Lagrange multipliers with the real auxilliary fields $\lambda_\nu$ 
 \begin{equation}
 Z_\phi=\int D\phi_sD\phi_cD\Phi_cD\Phi_sD\lambda_sD\lambda_c e^{-S_T} 
 \end{equation}
 with
 \begin{equation}
   S_T=S[\phi_c,\phi_s]+i\int_0^\beta d\tau \sum_\nu \left[ \Phi_\nu(\tau)-\phi_\nu(0,\tau) \right] \lambda_\nu(\tau) .
  \end{equation}
One needs consequently integrate out the $\phi_\nu$ and $\lambda_\nu$ to get the final baths actions:
 
\begin{equation}
S^\phi_\nu= \sum_{ \omega_n} |\omega_n| \left( \frac{  \hat{F}_{-\nu}}{\det({\cal F})} \right) |\Phi_\nu( \omega_n)|^2 
\end{equation}
with the bosonic Matsubara frequencies $\omega_n=\frac{2\pi n}{\beta} , (n \in \Z )$. The mixing term is:
\begin{equation}
S^\phi_{mix}=-\sum_{ \omega_n} |\omega_n|  \left( \frac{  \hat{F}_{mix}}{\det({\cal F})} \right) \Phi_s(-\omega_n) \Phi_c(\omega_n),
\label{SphiMix}
\end{equation}
where the ${\cal F}$ matrix and $\hat{F_\nu}$ are given in Eqs.~(\ref{FMat})-(\ref{Fmix}).  
 Importantly, the mixing action $S^\phi_{mix}$ vanishes (see Appendix) and one is left with the decoupled set:
  \begin{equation}
S^\Phi_\nu=\sum_{\omega_n}  \frac{|\omega_n| }{ \tilde{K}_\nu^\phi} |\Phi_\nu(\omega_n)|^2 
\label{EffectivePhiAction}
\end{equation}
with $\tilde{K}^\phi_\nu$ given by Eq.~(\ref{KPhiappendix}) and $\nu \in (s,c) $.
The resulted Caldeira-Legett type actions are common and describe dynamics of a single quantum Brownian particle in the regime of Ohmic dissipation~\cite{Caldeira1983,Zwerger1985}. These actions represent the weak-coupling fixed point of pure Luttinger liquid. 
 
Similarly, the effective baths actions for $\theta_\nu$ fields can be obtained as   
  \begin{equation}
S^\Theta_\nu= \sum_{\omega_n} |\omega_n| \tilde{K}^\theta_\nu | \Theta_\nu(\omega_n)|^2 
\label{EffectiveThetaAction}
\end{equation}
with the new TLL parameters $\tilde{K}_\nu^\theta$  given by Eq.~(\ref{KThetaappendix}). 
 The mixing action also vanishes in this case.
 
Vanishing mixing terms and decoupled baths map the problem onto the one with conventional TLL baths, but with the new parameters $\tilde{K_\nu}$ and $v_{-,+}$ given by Eqs.~(\ref{KPhiappendix})-(\ref{velNewModes}). Hence, the usual basis-rotation procedure is excessive here. Obviously, for the vanishing spin-charge mixing term $\Delta=0$, the standard TLL parameters are correctly recovered, $\tilde{K}_\nu \rightarrow K_\nu $. The characteristic velocities $ v_+ \rightarrow \max (v_s,v_c) $ and $v_- \rightarrow \min(v_s,v_c)$. This implies that in the presence of SOC, one has new modes, with carriers composed of spin and charge quanta. These new modes have excitation velocities $v_{-,+}$ and posses new TLL parameters $\tilde{K}_\nu$. 
 Prior to the discussion of our main results, we emphasize two important features arisen already at this stage of analysis.
 
 \textit{Mode freezing}. As the strength of SOC is increased, $v_+$ monotonically increases, whereas $v_-$ similarly decreases and vanishes at the critical $\Delta^c_\nu$. As it was mentioned in previous works~\cite{MorozTheory,Iucci}, at the critical SOC strength "freezing" of the corresponding mode (phase separation) with diverging spin (or charge) susceptibility is exhibited. For repulsive e-e interactions, it was shown~\cite{MorozTheory} that it is the spin susceptibility which diverges at the critical SOC as,   
   \begin{equation}
   \chi = \chi_0 \left( 1- \left(\frac{\Delta}{\Delta^c_s}\right)^2 \right)^{-1}
   \end{equation}
with spin-susceptibility $\chi_0$ at zero SOC and $\Delta_s^c=2/K_s$. Similar divergence can be observed in the charge channel, if one considers full $K_c-K_s$ space. These divergences indicate a phase transition, which occurs in the spin/charge channel (see Ref.~\cite{Iucci} and references therein). 
In this work, we consider SOC strengths causing the velocity difference $\Delta/v<0.8$, which restricts the parameter space to $K_\nu < 5/2$.

\textit{Absence of spin-filtering effect}. In the case of the spin-charge mixing caused by the strong Zeeman field [Fig.~\ref{fig:fig2}(a)] the mixing bath actions $S_{mix}$ does not vanish, due to the different type of SCC mechanism. The mixing term Eq.~(\ref{Hmixing}) in the Hamiltonian takes the following form:      
 \begin{equation}
H_{mix}=\frac{\Delta_B }{2} \int \left[\partial_x \phi_s \partial_x \phi_c + \partial_x \theta_s\partial_x \theta_c \right] dx 
\label{HmixingB}
\end{equation}
 with $\Delta_B=v_\uparrow-v_\downarrow$.
As it was demonstrated earlier in Refs.~\cite{KimuraResonantZeeman,Aoki1996}, a single impurity as in Eq.~(\ref{ImpHam}) embedded to such  TLL causes polarization of the spin current with a ratio of tunneling amplitudes, 
\begin{equation}
\frac{t_\uparrow}{t_\downarrow} = \left(\frac{T}{\Lambda}\right)^\eta,
\end{equation}  
and with a finite exponent $\eta(\Delta_B)$ for  finite $\Delta_B $.  
 
Observation of this effect in our model was also proposed earlier in Ref.~\cite{KimuraResonantZeeman}. However, an important consequence of vanishing $S_{mix}$ action terms of Eq.~(\ref{SphiMix}) (and similar term for $\Theta$ field) is the absence of such spin-filtering effect, albeit the spin-charge separations is violated.


  \section{Results and discussions}

  \label{sec:phasePortrait}
  \subsection*{Weak potential barrier  }
 Once the effective bath actions are obtained as in  Eq.~(\ref{EffectivePhiAction}) and Eq.~(\ref{EffectiveThetaAction}) , the standard pRG analysis in the limit of weak and strong impurity potential can be performed. 
 
We first consider the scattering of electrons on a weak barrier. In this limit, the partition function is, 

\begin{equation}
Z=\int D\Phi_s D\Phi_c e^{-S_{T}},
\label{EffectiveZ}
\end{equation}
with the total action, 
\begin{equation}
S_T = S^\Phi_c+S^\Phi_s + \frac{V_{m,n}}{\pi a} \int_0^\beta d\tau \cos(m \beta_I \Phi_c )\cos(n \beta_I \Phi_s ).
\label{TotalActPRG}
\end{equation}
To analyze the relevance of the last term within the $K_c-K_s$ parameter space, we generalize the impurity term to take different values of $m$ and $n$ with the amplitudes $V_{m,n}  \ll \Lambda $, since such terms are necessarily generated during the RG process.  

Treating the backscattering terms perturbatively, one obtains the standard set of first-order RG equations \cite{KaneImpurity2, NagaosaImpurity}, 
  \begin{equation}
 \frac{dV_{1,1}(l)}{dl}=\left( 1-\frac{\tilde{K}^\phi_s+\tilde{K}^\phi_c}{2} \right) V_{1,1}(l),
 \label{singleBS}
  \end{equation}
     \begin{equation}
 \frac{dV_{0,2}(l)}{dl}=\left( 1-2 \tilde{K}^\phi_s \right) V_{0,2}(l),
\label{spinBS}
  \end{equation}
      \begin{equation}
 \frac{dV_{2,0}(l)}{dl}=\left( 1-2 \tilde{K}^\phi_c \right) V_{2,0}(l)
 \label{chargeBS}
  \end{equation}
  with $dl = \frac{d\Lambda}{\Lambda}$.

\begin{figure}[t]
\includegraphics[width=8.5cm ]{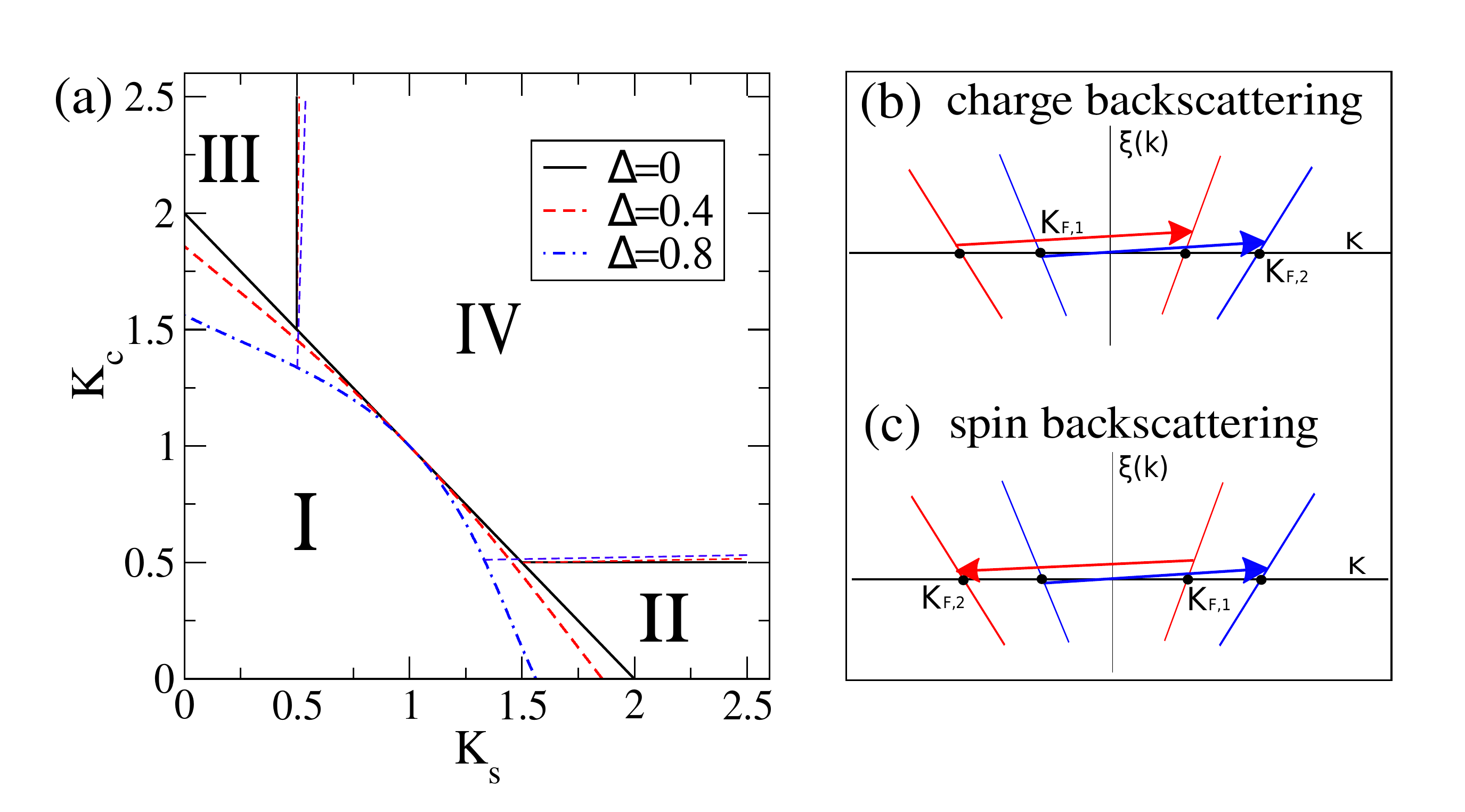}%
\caption{\label{fig:fig3} (a) The modification of the phase portrait for the TLL with finite SOC. For $\Delta>0$ the phase boundary of region I is modified. In (b) and (c) the backscattering processes of charge and spin are sketched, which correspond to Eqs.~(\ref{spinBS})-(\ref{chargeBS}), respectively. 
} 
\end{figure}

   The first equation for $V_{1,1}  \equiv V_{bs}$ corresponds to the $k_{F,1}+k_{F,2}$ backscattering of a single electron, whereas the second and the third equations correspond to the backscattering of spin or charge (electron pair) degree of freedom. The last processes have the fermionic expressions  $\psi^\dagger_{R,\uparrow}\psi_{L,\uparrow}\psi^\dagger_{L,\downarrow}\psi_{R,\downarrow} +h.c.$ and $\psi^\dagger_{R,\uparrow}\psi_{L,\uparrow}\psi^\dagger_{R,\downarrow}\psi_{L,\downarrow} +h.c.$, respectively. The sketches of these scattering processes in the momentum space are presented in Fig.~\ref{fig:fig3}(b)-(c). All higher-order decendent terms are neglected, since the regions of their relevancy are covered with the ones of the last two equations, even for the finite $\Delta$. Thus, the main low-temperature  processes are dictated by these three equations~Eqs.(\ref{singleBS})-(\ref{chargeBS}).
   

%
%
For vanishing SOC $\Delta=0$ ($\tilde{K}_\nu \rightarrow K_\nu$), the marginal fixed line defined in Eq.~(\ref{singleBS}) is determined by the condition $K_s+K_c=2$. The region I in Fig.~\ref{fig:fig3}(a) corresponds to a parameter space where the backscattering of a single electron in a weak potential becomes a relevant process and the backscattering amplitude $V_{1,1}$ flows to the strong-coupling regime. This leads to the blocked transport in both charge and spin channels and the fields $\Phi_{s,c}$ are pinned on the minima of cosine functions. Qualitatively, in this region either single electron $(\uparrow $ or $ \downarrow)$  is \textit{fractionalized} by e-e interactions, which is responsible for full backscattering of single   carrier and hence, charge and spin carriers, too. Thus, one has vanishing conductance  $G =0$ at $T=0$.
	
	As dictated by Eqs.~(\ref{spinBS})-(\ref{chargeBS}), the backscattering processes of the charge/spin carriers also become relevant when $K_{c/s}<0.5$. For the charge backscattering, this implies the charge quanta to carry smaller than $0.5$ unit, to make it a relevant process. In the region of relevance II (III for spin channel), the charge (spin) channel is insulating since the the $\Phi_c$ ($\Phi_s$) is pinned. The spin (charge) fully transmit the barrier in this region, implying the realization of mixed phases in these regions.  
	  In region IV both channels have dominating superconducting fluctuations~\cite{Iucci}. We emphasize again that in this region with attractive e-e interactions, other type of instabilities in the bulk may arise, which usually open a gap in one of the channels. Here, we neglect all such processes and present the simplest picture. 

 To investigate the effect of SOC on backscattering of carriers, we represent the new TLL parameters in a symmetric way: 
	\begin{equation}
	\tilde{K}^\phi_\nu=K_\nu \left(\frac{v_\nu}{\tilde{v}_\nu} \right) \left(\frac{\tilde{v}_c+\tilde{v}_s}{v_++v_-}\right),
	\label{newKphi}
    \end{equation}	  
 where $\tilde{v}_\nu$  and $v_{\pm}$ are given in Eq.~(\ref{veltildeDef}) and Eq.~(\ref{velNewModes}), respectively.
 In the presence of SOC the first effect is the renormalization of spin and charge velocities, which is manifested in the first factor of Eq.~(\ref{newKphi}). The second factor is due to the emerged modes with velocities $v_{+,-}$.   As it was mentioned in the previous section, we limit considiration of the parameter space $K_\nu<\frac{5}{2}$ and $\Delta/v <0.8$.

For non-interacting electrons, the effect of SOC is limited to the breaking of chiral symmetry and the renormalization of excitation velocities. Indeed, one has $\tilde{K}_\nu^{\phi}=K_\nu^\phi=1$, since renormalizing factors in Eq.~(\ref{newKphi}) cancel each other. As shown in Fig.~\ref{fig:fig3}a this is also valid for weakly-interacting electrons, $K_\nu \sim 1$.

For the given finite SOC strengths, where $ 0<\Delta/v<0.8$, the second  factor  $\left(\frac{\tilde{v}_s+\tilde{v}_c}{v_-+v_+} \right) \approx  1$ is fixed  for both $K_\nu$, whereas the first one defines scattering of carriers and determines the boundary of region I.  
The strongest effect of SOC is exhibited when in one of the  channels $K_\nu\ll 1$. In this limit one can lock the corresponding $\Phi_\nu$ field to the minimum of cosine function and reexpress the scaling dimension of the impurity term for $m=1$ and $n=1$ as follows: 
\begin{equation}
2\delta_{1,1}=  K_\nu+K_{-\nu} (1+\frac{\Delta^2}{8v_{-\nu}^2} ). 
\end{equation}
  The resulting marginality line is presented in Fig.~\ref{fig:fig3}(a) for $\Delta/v=0.4$ and $\Delta/v=0.8$. At the critical $K_\nu=\frac{5}{2}$, the excitations in the spin/charge channel become frozen ($v_-=0$) and the bulk of the channel becomes insulating. The effect of finite SOC on the boundaries of regions II/III with the region IV is negligibally small. The largest correction to the scaling dimension $\delta_{0,1} (\delta_{1,0}$) is of the order of $10^{-2}$ for the largest $\Delta/v=0.8$. Thereby it can be safely neglected.  
 
  Finally, one can straightforwardly generalize the expressions for corrections to (bulk) conductances obtained in Refs.~\cite{NagaosaImpurity}\cite{KaneImpurity2} to the case with finite SOC by $K_\nu \rightarrow \tilde{K}_\nu$:
  
 \begin{equation}
 \delta G=\frac{e^2}{h}\sum_{m,n} c_{m,n} |V_{m,n}|^2 T^{\left( m^2 \tilde{K}_c + n^2 \tilde{K}_s \right)/2-2}  
 \end{equation}
   with dimensionless coefficients $c_{m,n}$. 
    \subsection*{Strong barrier }
    For analysis of tunneling term in Eq.~(\ref{TunnelingImp}) one considers the following total action $S_T$,
 \begin{equation}
S_T = S^\Theta_c+S^\Theta_s + \frac{t_{m,n}}{\pi a} \int_0^\beta d\tau \cos(m \beta_I \Theta_c )\cos(n \beta_I \Theta_s ).
\label{TotalActPRG1}
\end{equation}
Similar to the previous case, the impurity term is generalized for different $m$ and $n$.

The RG transformation of the impurity term leads to the following set of equations, 
 
\begin{equation}
\frac{dt_{1,1}(l)}{dl}=\left( 1-\frac{1}{2}\left( \frac{1}{\tilde{K}_c^\theta}+\frac{1}{\tilde{K}_s^\theta} \right) \right)  t_{1,1}(l),
\label{singleTunn}
\end{equation}
 \begin{equation}
\frac{dt_{2,0}(l)}{dl}=\left( 1-\frac{2}{\tilde{K}_c^\theta}  \right)  t_{2,0}(l),
\label{chargeTunn}
\end{equation}
  \begin{equation}
\frac{dt_{0,2}(l)}{dl}=\left( 1-\frac{2}{\tilde{K}_s^\theta}  \right)  t_{0,2}(l) 
\label{spinTunn}
\end{equation}
 with the TLL prameters rewritten as: 
\begin{equation}
\frac{1}{\tilde{K}_\nu^\theta}=\frac{1}{K_\nu}\left( \frac{v_\nu}{\hat{v}_\nu} \right) \left( \frac{\hat{v}_c+\hat{v}_s}{v_-+v_+}\right),
 \end{equation} 
and
\begin{equation}
\hat{v}_\nu=\frac{v_\nu}{\sqrt{1-\frac{\Delta^2}{4v_\nu^2}}}.
\end{equation}
   \begin{figure}[t]
\includegraphics[width=8.5cm ]{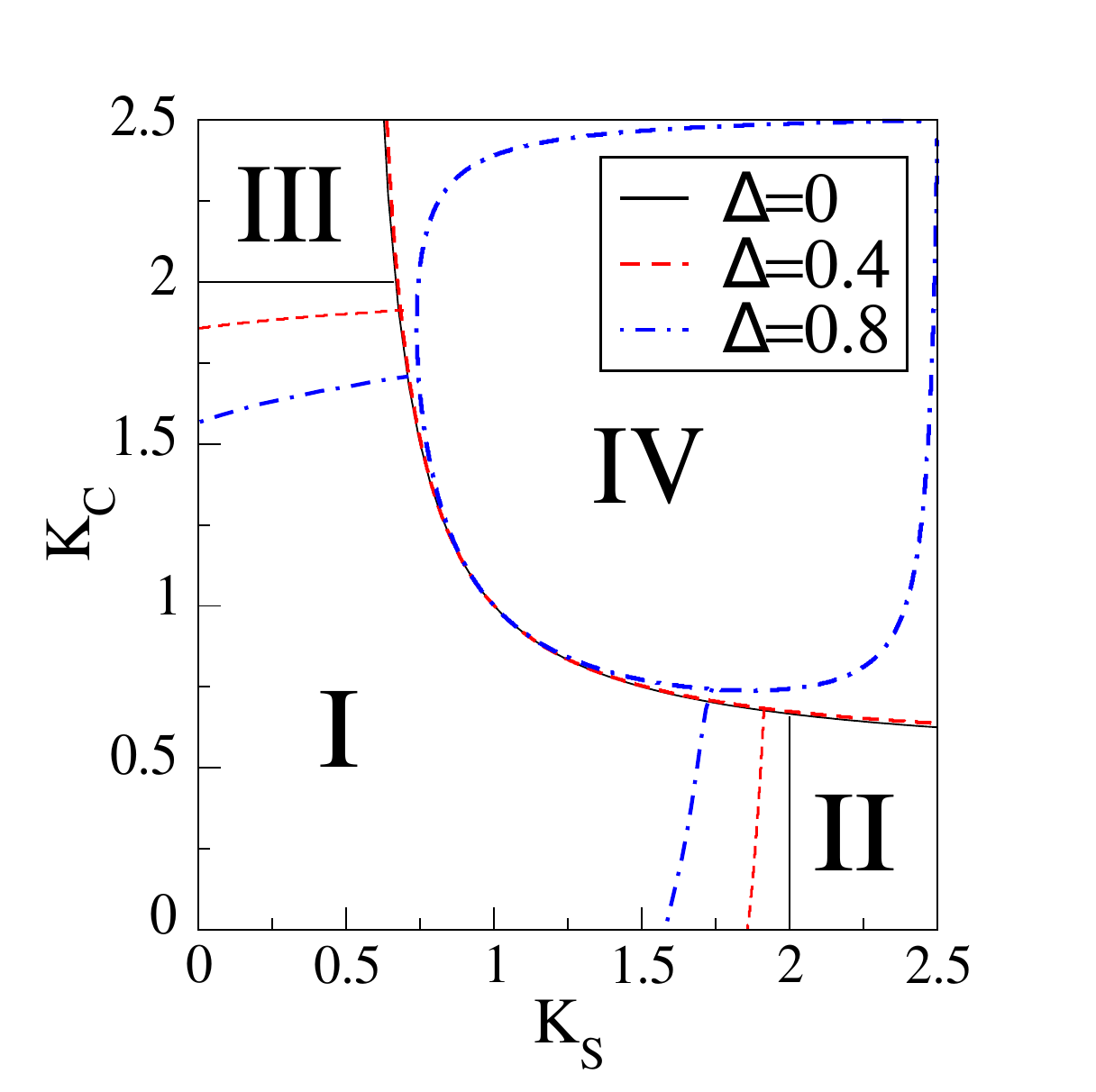}%
\caption{\label{fig:fig4} (a) The modification of the phase portrait obtained from the renormalization analysis of tunneling events. The effect of SOC is manifested on phase boundaries between the regions I-II and I-III. In the vicinity of mode freezing, hopping events of a single $(\uparrow)/(\downarrow$) carrier become irrelevant. All shaded regions correspond to disconnected wires with no transport of carriers.   
} 
\end{figure}
 
An amplidute $t_{1,1}$  corresponds  to the interwire hopping event of a single $(\uparrow)/(\downarrow)$ electron, while $t_{2,0}$ and $t_{0,2}$ are tunneling amplitude of charge and spin, respectively.

 Based on the set of equations, one obtains phase portrait at $T=0$. It consists of four regions, exhibited also in the limit of weak scattering potential and it is shown in Fig.~\ref{fig:fig4}(a). 
  For vanishing SOC $\tilde{K_\nu} \rightarrow K_\nu$, the boundaries of regions are defined by $ \frac{1}{K_c}+\frac{1}{K_s}=2,  K_c=2$, and  $ K_s=2$.
 In region I (IV), hopping event of a single $(\uparrow)/(\downarrow)$ electron is irrelevant (relevant) and one eventually renormalizes onto the fixed point of disjoined (connected) wires in the RG process. In region II, the tunneling amplitude $t_{0,2}$ for spin is relevant and grows upon RG transformation, however charge carriers can not tunnel. An opposite situation with a conducting charge channel and an insulating spin channel occurs in the region III. We note that the results obtained in the opposite limits of weak and strong barrier are consistent.   
 
 For relatively weak SOC ($\Delta=0.4$) the only pronounced effect is modification of boundaries between regions I-II and I-III, as shown in Fig.~\ref{fig:fig4}(a). These effects are dictated by Eqs.~(\ref{chargeTunn})-(\ref{spinTunn}). The regions II and III are extended towards the region I with the new boundaries  $K_\nu \approx 1.8 <2$. The area of the extended region is larger for the larger value of $\Delta=0.8$ with boundaries defined via $K_\nu \approx 1.6$. Remarkably, these effects are also exhibited in the weak barrier analysis, in Fig.~\ref{fig:fig3}(a).

 At the largest $\Delta=0.8$, the tunneling of single carriers with amplitude $t_{1,1}$ becomes irrelevant near the phase separation point $K_\nu=5/2$. This is consistent with the picture of mode freezing, discussed in the previous sections. On the other hand, for weakly-interacting electrons with $K_\nu \approx 1$ the effect of SOC is negligibely small, even for large values of $\Delta$, also compatible with the results of the previous section.
 
 Conductance for insulating regions at finite temperature $T$ can be also generalized as follows:
 \begin{equation}
 G=\frac{e^2}{h} \sum_{m,n} d_{m,n} t_{m,n}^2 T^{2\left(m^2/\tilde{K}_c+n^2/\tilde{K}_s -2\right)} 
 \end{equation}
 with dimensionless coefficients $d_{m,n}$.
  
 Experimentally relevant values of $\Delta/v \sim 0.2$ \cite{MorozTheory,MorozExpr,SatoExperiment} are smaller than the maximum value considered in this work. For this range of $\Delta$ values, one can neglect mode-freezing effects safely. The results of the RG analysis on the last sections affirm that for weak and moderate e-e interactions  effects of SOC are negligible small. This results are in accrodance with recent theoretical and expertimental studies \cite{SatoExperiment,SatoTheory}.  
  \section{Conclusions}
  We studied carrier scattering effects upon a single impurity embedded to a Luttinger liquid with spin-orbit coupling using Abelian bosonization and pertubative renormalization techniques. Spin-orbit interaction degrades spin-charge separation and renormalizes the TLL parameters $K_{s,c}$ and the excitation velocities $v_{s,c}$. We demonstrated that the scaling dimension of impurity operator is identical for both $(\uparrow)/(\downarrow)$ carriers. This implies the absence of conjectured spin-filtering effect in Ref.\cite{KimuraResonantZeeman}. The strongest effects of spin-orbit coupling are pronounced for strong e-e interactions, whereas these effects are negligibly small for moderate e-e interactions. Our main results are summarized by the phase portrait modifications presented in Figs~(\ref{fig:fig3})-(\ref{fig:fig4}).

\begin{acknowledgments}
 We thank F. Yilmaz and V.~I. Yudson for useful comments and discussions. 
\end{acknowledgments}
\newpage
\appendix*
\section{ Expressions for new TLL parameters and excitations velocities}
 The ${\cal F}$ matrix has the following form,  
\begin{equation}
{\cal F}=\begin{pmatrix}
2\hat{F}_c& \hat{F}_{mix}\\
\hat{F}_{mix} & 2\hat{F}_s\\
\end{pmatrix}.
\label{FMat}
\end{equation}
The matrix elements $\hat{F_\nu}=|\omega|F_\nu$ are expressed via the following integrals, 
 \begin{equation}
F_\nu=\int_{-\infty} ^{+\infty} \frac{dk}{2\pi}\left( \frac{[{\cal{G}}_{-\nu}^\phi]^{-1}}{\det({\cal Q})} \right)
\label{Fs}
\end{equation}
 
and for the mixing element,
\begin{equation}
F_{mix}=-\int_{-\infty} ^{+\infty}  \frac{dk}{2\pi} \left( \frac{[{\cal{G}}_{mix}^\phi]^{-1}}{\det({\cal Q})} \right)
\label{Fmix}
\end{equation}
 where $[{\cal{G}}_\nu^\phi]^{-1}$ are the inverse propogators of Eqs.~(\ref{actionPhiFull})-(\ref{actionPhimix}) in $\vec{q}=(k,\omega)$ space,  and the ${\cal Q}$ matrix is defined as,
\begin{equation}
{\cal Q}=\begin{pmatrix}
2[{\cal{G}}_{c}^\phi]^{-1}& [{\cal{G}}_{mix}^\phi]^{-1}\\
[{\cal{G}}_{mix}^\phi]^{-1} & 2[{\cal{G}}_{s}^\phi]^{-1}\\
\end{pmatrix}.
\end{equation} 
 
 For the mixing parameter $F_{mix}$ the numerator of the kernel is odd function of $k$, i.e $[{\cal{G}}_{mix}^\phi]^{-1}=k\omega $, whereas the denominator is an even function of $k$. This leads to the vanishing action $S^\phi_{mix}$. Similarly, $S^\theta_{mix}$ vanishes and one is left with fully decoupled actions [Eqs.~(\ref{EffectivePhiAction})-(\ref{EffectiveThetaAction})].
 
Evaluation of integrals leads to the following results for new TLL parameters for displacement fluctuation fields, 
\begin{equation}
\hat{K}_\nu^{\phi}= \frac{v}{ (v_++v_-)}\left[ \frac{\tilde{v}_{-\nu}^2}{v_+v_-} +1 \right],
\label{KPhiappendix}
\end{equation}

and for the phase fields, 
 \begin{equation}
\frac{1}{\hat{K}_\nu^{\theta}}= \frac{v_\nu^2}{ v( v_- + v_+  )} \left[  \frac{v_{-\nu}^2 \tilde{v}_\nu}{  v_\nu^2\tilde{v}_{-\nu}} +1 \right].
\label{KThetaappendix}
\end{equation}
 
The sound velocities for the newly emerged modes are given by, 
 
\begin{equation}
v^2_\pm=\frac{1}{2} \left[ \tilde{v}_c^2+\tilde{v}_s^2+\Delta^2\pm \sqrt{(\tilde{v}_c^2-\tilde{v}_s^2+\Delta^2)^2+4\tilde{v}_s^2\Delta^2} \right].
\label{velNewModes}
\end{equation}

For vanishing SOC $\Delta=0$, the sound velocities in the spin/charge channel are recovered.
 
   \bibliographystyle{apsrev4-1}
 \bibliography{refs}

\begin{thebibliography}{43}%
\makeatletter
\providecommand \@ifxundefined [1]{%
 \@ifx{#1\undefined}
}%
\providecommand \@ifnum [1]{%
 \ifnum #1\expandafter \@firstoftwo
 \else \expandafter \@secondoftwo
 \fi
}%
\providecommand \@ifx [1]{%
 \ifx #1\expandafter \@firstoftwo
 \else \expandafter \@secondoftwo
 \fi
}%
\providecommand \natexlab [1]{#1}%
\providecommand \enquote  [1]{``#1''}%
\providecommand \bibnamefont  [1]{#1}%
\providecommand \bibfnamefont [1]{#1}%
\providecommand \citenamefont [1]{#1}%
\providecommand \href@noop [0]{\@secondoftwo}%
\providecommand \href [0]{\begingroup \@sanitize@url \@href}%
\providecommand \@href[1]{\@@startlink{#1}\@@href}%
\providecommand \@@href[1]{\endgroup#1\@@endlink}%
\providecommand \@sanitize@url [0]{\catcode `\\12\catcode `\$12\catcode
  `\&12\catcode `\#12\catcode `\^12\catcode `\_12\catcode `\%12\relax}%
\providecommand \@@startlink[1]{}%
\providecommand \@@endlink[0]{}%
\providecommand \url  [0]{\begingroup\@sanitize@url \@url }%
\providecommand \@url [1]{\endgroup\@href {#1}{\urlprefix }}%
\providecommand \urlprefix  [0]{URL }%
\providecommand \Eprint [0]{\href }%
\providecommand \doibase [0]{http://dx.doi.org/}%
\providecommand \selectlanguage [0]{\@gobble}%
\providecommand \bibinfo  [0]{\@secondoftwo}%
\providecommand \bibfield  [0]{\@secondoftwo}%
\providecommand \translation [1]{[#1]}%
\providecommand \BibitemOpen [0]{}%
\providecommand \bibitemStop [0]{}%
\providecommand \bibitemNoStop [0]{.\EOS\space}%
\providecommand \EOS [0]{\spacefactor3000\relax}%
\providecommand \BibitemShut  [1]{\csname bibitem#1\endcsname}%
\let\auto@bib@innerbib\@empty
\bibitem [{\citenamefont {Tomonaga}(1950)}]{TomonagaTLL}%
  \BibitemOpen
  \bibfield  {author} {\bibinfo {author} {\bibfnamefont {S.-i.}\ \bibnamefont
  {Tomonaga}},\ }\href@noop {} {\bibfield  {journal} {\bibinfo  {journal}
  {Progress of Theoretical Physics}\ }\textbf {\bibinfo {volume} {5}},\
  \bibinfo {pages} {544} (\bibinfo {year} {1950})}\BibitemShut {NoStop}%
\bibitem [{\citenamefont {Luttinger}(1963)}]{LuttingerTLL}%
  \BibitemOpen
  \bibfield  {author} {\bibinfo {author} {\bibfnamefont {J.}~\bibnamefont
  {Luttinger}},\ }\href@noop {} {\bibfield  {journal} {\bibinfo  {journal}
  {Journal of mathematical physics}\ }\textbf {\bibinfo {volume} {4}},\
  \bibinfo {pages} {1154} (\bibinfo {year} {1963})}\BibitemShut {NoStop}%
\bibitem [{\citenamefont {Gogolin}\ \emph {et~al.}(2004)\citenamefont
  {Gogolin}, \citenamefont {Nersesyan},\ and\ \citenamefont
  {Tsvelik}}]{GogolinTLL}%
  \BibitemOpen
  \bibfield  {author} {\bibinfo {author} {\bibfnamefont {A.~O.}\ \bibnamefont
  {Gogolin}}, \bibinfo {author} {\bibfnamefont {A.~A.}\ \bibnamefont
  {Nersesyan}}, \ and\ \bibinfo {author} {\bibfnamefont {A.~M.}\ \bibnamefont
  {Tsvelik}},\ }\href@noop {} {\emph {\bibinfo {title} {Bosonization and
  strongly correlated systems}}}\ (\bibinfo  {publisher} {Cambridge university
  press},\ \bibinfo {year} {2004})\BibitemShut {NoStop}%
\bibitem [{\citenamefont {Haldane}(1981)}]{HaldaneTLL}%
  \BibitemOpen
  \bibfield  {author} {\bibinfo {author} {\bibfnamefont {F.~D.~M.}\
  \bibnamefont {Haldane}},\ }\href {\doibase 10.1088/0022-3719/14/19/010} {\
  \textbf {\bibinfo {volume} {14}},\ \bibinfo {pages} {2585} (\bibinfo {year}
  {1981})}\BibitemShut {NoStop}%
\bibitem [{\citenamefont {Fisher}\ and\ \citenamefont
  {Glazman}(1997)}]{Glazman}%
  \BibitemOpen
  \bibfield  {author} {\bibinfo {author} {\bibfnamefont {M.~P.}\ \bibnamefont
  {Fisher}}\ and\ \bibinfo {author} {\bibfnamefont {L.~I.}\ \bibnamefont
  {Glazman}},\ }in\ \href@noop {} {\emph {\bibinfo {booktitle} {Mesoscopic
  Electron Transport}}}\ (\bibinfo  {publisher} {Springer},\ \bibinfo {year}
  {1997})\ pp.\ \bibinfo {pages} {331--373}\BibitemShut {NoStop}%
\bibitem [{\citenamefont {Steinberg}\ \emph {et~al.}(2008)\citenamefont
  {Steinberg}, \citenamefont {Barak}, \citenamefont {Yacoby}, \citenamefont
  {Pfeiffer}, \citenamefont {West}, \citenamefont {Halperin},\ and\
  \citenamefont {Le~Hur}}]{KLeHurFrac}%
  \BibitemOpen
  \bibfield  {author} {\bibinfo {author} {\bibfnamefont {H.}~\bibnamefont
  {Steinberg}}, \bibinfo {author} {\bibfnamefont {G.}~\bibnamefont {Barak}},
  \bibinfo {author} {\bibfnamefont {A.}~\bibnamefont {Yacoby}}, \bibinfo
  {author} {\bibfnamefont {L.~N.}\ \bibnamefont {Pfeiffer}}, \bibinfo {author}
  {\bibfnamefont {K.~W.}\ \bibnamefont {West}}, \bibinfo {author}
  {\bibfnamefont {B.~I.}\ \bibnamefont {Halperin}}, \ and\ \bibinfo {author}
  {\bibfnamefont {K.}~\bibnamefont {Le~Hur}},\ }\href {\doibase
  10.1038/nphys810} {\bibfield  {journal} {\bibinfo  {journal} {Nature
  Physics}\ }\textbf {\bibinfo {volume} {4}},\ \bibinfo {pages} {116} (\bibinfo
  {year} {2008})}\BibitemShut {NoStop}%
\bibitem [{\citenamefont {Kamata}\ \emph {et~al.}(2014)\citenamefont {Kamata},
  \citenamefont {Kumada}, \citenamefont {Hashisaka}, \citenamefont {Muraki},\
  and\ \citenamefont {Fujisawa}}]{KamataFrac}%
  \BibitemOpen
  \bibfield  {author} {\bibinfo {author} {\bibfnamefont {H.}~\bibnamefont
  {Kamata}}, \bibinfo {author} {\bibfnamefont {N.}~\bibnamefont {Kumada}},
  \bibinfo {author} {\bibfnamefont {M.}~\bibnamefont {Hashisaka}}, \bibinfo
  {author} {\bibfnamefont {K.}~\bibnamefont {Muraki}}, \ and\ \bibinfo {author}
  {\bibfnamefont {T.}~\bibnamefont {Fujisawa}},\ }\href {\doibase
  10.1038/nnano.2013.312} {\bibfield  {journal} {\bibinfo  {journal} {Nature
  Nanotechnology}\ }\textbf {\bibinfo {volume} {9}},\ \bibinfo {pages} {177}
  (\bibinfo {year} {2014})}\BibitemShut {NoStop}%
\bibitem [{\citenamefont {Voit}(1995)}]{VoitTLL}%
  \BibitemOpen
  \bibfield  {author} {\bibinfo {author} {\bibfnamefont {J.}~\bibnamefont
  {Voit}},\ }\href@noop {} {\bibfield  {journal} {\bibinfo  {journal} {Reports
  on Progress in Physics}\ }\textbf {\bibinfo {volume} {58}},\ \bibinfo {pages}
  {977} (\bibinfo {year} {1995})}\BibitemShut {NoStop}%
\bibitem [{\citenamefont {Giamarchi}(2003)}]{GiamarchiTLL}%
  \BibitemOpen
  \bibfield  {author} {\bibinfo {author} {\bibfnamefont {T.}~\bibnamefont
  {Giamarchi}},\ }\href@noop {} {\emph {\bibinfo {title} {Quantum physics in
  one dimension}}},\ Vol.\ \bibinfo {volume} {121}\ (\bibinfo  {publisher}
  {Clarendon press},\ \bibinfo {year} {2003})\BibitemShut {NoStop}%
\bibitem [{\citenamefont {Jeckelmann}(2012)}]{JeckelmanDOS}%
  \BibitemOpen
  \bibfield  {author} {\bibinfo {author} {\bibfnamefont {E.}~\bibnamefont
  {Jeckelmann}},\ }\href {\doibase 10.1088/0953-8984/25/1/014002} {\bibfield
  {journal} {\bibinfo  {journal} {Journal of Physics: Condensed Matter}\
  }\textbf {\bibinfo {volume} {25}},\ \bibinfo {pages} {014002} (\bibinfo
  {year} {2012})}\BibitemShut {NoStop}%
\bibitem [{\citenamefont {Kane}\ and\ \citenamefont
  {Fisher}(1992{\natexlab{a}})}]{KaneImpurity}%
  \BibitemOpen
  \bibfield  {author} {\bibinfo {author} {\bibfnamefont {C.}~\bibnamefont
  {Kane}}\ and\ \bibinfo {author} {\bibfnamefont {M.~P.}\ \bibnamefont
  {Fisher}},\ }\href@noop {} {\bibfield  {journal} {\bibinfo  {journal} {Phys.
  Rev. Lett.}\ }\textbf {\bibinfo {volume} {68}},\ \bibinfo {pages} {1220}
  (\bibinfo {year} {1992}{\natexlab{a}})}\BibitemShut {NoStop}%
\bibitem [{\citenamefont {Furusaki}\ and\ \citenamefont
  {Nagaosa}(1993)}]{NagaosaImpurity}%
  \BibitemOpen
  \bibfield  {author} {\bibinfo {author} {\bibfnamefont {A.}~\bibnamefont
  {Furusaki}}\ and\ \bibinfo {author} {\bibfnamefont {N.}~\bibnamefont
  {Nagaosa}},\ }\href@noop {} {\bibfield  {journal} {\bibinfo  {journal} {Phys.
  Rev. B}\ }\textbf {\bibinfo {volume} {47}},\ \bibinfo {pages} {4631}
  (\bibinfo {year} {1993})}\BibitemShut {NoStop}%
\bibitem [{\citenamefont {Kane}\ and\ \citenamefont
  {Fisher}(1992{\natexlab{b}})}]{KaneImpurity2}%
  \BibitemOpen
  \bibfield  {author} {\bibinfo {author} {\bibfnamefont {C.}~\bibnamefont
  {Kane}}\ and\ \bibinfo {author} {\bibfnamefont {M.~P.}\ \bibnamefont
  {Fisher}},\ }\href@noop {} {\bibfield  {journal} {\bibinfo  {journal} {Phys.
  Rev. B}\ }\textbf {\bibinfo {volume} {46}},\ \bibinfo {pages} {15233}
  (\bibinfo {year} {1992}{\natexlab{b}})}\BibitemShut {NoStop}%
\bibitem [{\citenamefont {Anthore}\ \emph {et~al.}(2018)\citenamefont
  {Anthore}, \citenamefont {Iftikhar}, \citenamefont {Boulat}, \citenamefont
  {Parmentier}, \citenamefont {Cavanna}, \citenamefont {Ouerghi}, \citenamefont
  {Gennser},\ and\ \citenamefont {Pierre}}]{QuantumSimulator}%
  \BibitemOpen
  \bibfield  {author} {\bibinfo {author} {\bibfnamefont {A.}~\bibnamefont
  {Anthore}}, \bibinfo {author} {\bibfnamefont {Z.}~\bibnamefont {Iftikhar}},
  \bibinfo {author} {\bibfnamefont {E.}~\bibnamefont {Boulat}}, \bibinfo
  {author} {\bibfnamefont {F.~D.}\ \bibnamefont {Parmentier}}, \bibinfo
  {author} {\bibfnamefont {A.}~\bibnamefont {Cavanna}}, \bibinfo {author}
  {\bibfnamefont {A.}~\bibnamefont {Ouerghi}}, \bibinfo {author} {\bibfnamefont
  {U.}~\bibnamefont {Gennser}}, \ and\ \bibinfo {author} {\bibfnamefont
  {F.}~\bibnamefont {Pierre}},\ }\href {\doibase 10.1103/PhysRevX.8.031075}
  {\bibfield  {journal} {\bibinfo  {journal} {Phys. Rev. X}\ }\textbf {\bibinfo
  {volume} {8}},\ \bibinfo {pages} {031075} (\bibinfo {year}
  {2018})}\BibitemShut {NoStop}%
\bibitem [{\citenamefont {Chang}(2003)}]{RealisticTLLChiralQHE}%
  \BibitemOpen
  \bibfield  {author} {\bibinfo {author} {\bibfnamefont {A.~M.}\ \bibnamefont
  {Chang}},\ }\href {\doibase 10.1103/RevModPhys.75.1449} {\bibfield  {journal}
  {\bibinfo  {journal} {Rev. Mod. Phys.}\ }\textbf {\bibinfo {volume} {75}},\
  \bibinfo {pages} {1449} (\bibinfo {year} {2003})}\BibitemShut {NoStop}%
\bibitem [{\citenamefont {Wen}(1990)}]{RealisticTLLChiralQHE2}%
  \BibitemOpen
  \bibfield  {author} {\bibinfo {author} {\bibfnamefont {X.~G.}\ \bibnamefont
  {Wen}},\ }\href {\doibase 10.1103/PhysRevB.41.12838} {\bibfield  {journal}
  {\bibinfo  {journal} {Phys. Rev. B}\ }\textbf {\bibinfo {volume} {41}},\
  \bibinfo {pages} {12838} (\bibinfo {year} {1990})}\BibitemShut {NoStop}%
\bibitem [{\citenamefont {Bockrath}\ \emph {et~al.}(1999)\citenamefont
  {Bockrath}, \citenamefont {Cobden}, \citenamefont {Lu}, \citenamefont
  {Rinzler}, \citenamefont {Smalley}, \citenamefont {Balents},\ and\
  \citenamefont {McEuen}}]{RealisticTLLNanotube}%
  \BibitemOpen
  \bibfield  {author} {\bibinfo {author} {\bibfnamefont {M.}~\bibnamefont
  {Bockrath}}, \bibinfo {author} {\bibfnamefont {D.~H.}\ \bibnamefont
  {Cobden}}, \bibinfo {author} {\bibfnamefont {J.}~\bibnamefont {Lu}}, \bibinfo
  {author} {\bibfnamefont {A.~G.}\ \bibnamefont {Rinzler}}, \bibinfo {author}
  {\bibfnamefont {R.~E.}\ \bibnamefont {Smalley}}, \bibinfo {author}
  {\bibfnamefont {L.}~\bibnamefont {Balents}}, \ and\ \bibinfo {author}
  {\bibfnamefont {P.~L.}\ \bibnamefont {McEuen}},\ }\href {\doibase
  10.1038/17569} {\bibfield  {journal} {\bibinfo  {journal} {Nature}\ }\textbf
  {\bibinfo {volume} {397}},\ \bibinfo {pages} {598} (\bibinfo {year}
  {1999})}\BibitemShut {NoStop}%
\bibitem [{\citenamefont {Blumenstein}\ \emph {et~al.}(2011)\citenamefont
  {Blumenstein}, \citenamefont {Sch{\"a}fer}, \citenamefont {Mietke},
  \citenamefont {Meyer}, \citenamefont {Dollinger}, \citenamefont {Lochner},
  \citenamefont {Cui}, \citenamefont {Patthey}, \citenamefont {Matzdorf},\ and\
  \citenamefont {Claessen}}]{RealisticTLLWire}%
  \BibitemOpen
  \bibfield  {author} {\bibinfo {author} {\bibfnamefont {C.}~\bibnamefont
  {Blumenstein}}, \bibinfo {author} {\bibfnamefont {J.}~\bibnamefont
  {Sch{\"a}fer}}, \bibinfo {author} {\bibfnamefont {S.}~\bibnamefont {Mietke}},
  \bibinfo {author} {\bibfnamefont {S.}~\bibnamefont {Meyer}}, \bibinfo
  {author} {\bibfnamefont {A.}~\bibnamefont {Dollinger}}, \bibinfo {author}
  {\bibfnamefont {M.}~\bibnamefont {Lochner}}, \bibinfo {author} {\bibfnamefont
  {X.~Y.}\ \bibnamefont {Cui}}, \bibinfo {author} {\bibfnamefont
  {L.}~\bibnamefont {Patthey}}, \bibinfo {author} {\bibfnamefont
  {R.}~\bibnamefont {Matzdorf}}, \ and\ \bibinfo {author} {\bibfnamefont
  {R.}~\bibnamefont {Claessen}},\ }\href {\doibase 10.1038/nphys2051}
  {\bibfield  {journal} {\bibinfo  {journal} {Nature Physics}\ }\textbf
  {\bibinfo {volume} {7}},\ \bibinfo {pages} {776} (\bibinfo {year}
  {2011})}\BibitemShut {NoStop}%
\bibitem [{\citenamefont {Jompol}\ \emph {et~al.}(2009)\citenamefont {Jompol},
  \citenamefont {Ford}, \citenamefont {Griffiths}, \citenamefont {Farrer},
  \citenamefont {Jones}, \citenamefont {Anderson}, \citenamefont {Ritchie},
  \citenamefont {Silk},\ and\ \citenamefont {Schofield}}]{JompolSCS}%
  \BibitemOpen
  \bibfield  {author} {\bibinfo {author} {\bibfnamefont {Y.}~\bibnamefont
  {Jompol}}, \bibinfo {author} {\bibfnamefont {C.~J.~B.}\ \bibnamefont {Ford}},
  \bibinfo {author} {\bibfnamefont {J.~P.}\ \bibnamefont {Griffiths}}, \bibinfo
  {author} {\bibfnamefont {I.}~\bibnamefont {Farrer}}, \bibinfo {author}
  {\bibfnamefont {G.~A.~C.}\ \bibnamefont {Jones}}, \bibinfo {author}
  {\bibfnamefont {D.}~\bibnamefont {Anderson}}, \bibinfo {author}
  {\bibfnamefont {D.~A.}\ \bibnamefont {Ritchie}}, \bibinfo {author}
  {\bibfnamefont {T.~W.}\ \bibnamefont {Silk}}, \ and\ \bibinfo {author}
  {\bibfnamefont {A.~J.}\ \bibnamefont {Schofield}},\ }\href {\doibase
  10.1126/science.1171769} {\bibfield  {journal} {\bibinfo  {journal}
  {Science}\ }\textbf {\bibinfo {volume} {325}},\ \bibinfo {pages} {597}
  (\bibinfo {year} {2009})}\BibitemShut {NoStop}%
\bibitem [{\citenamefont {Bocquillon}\ \emph {et~al.}(2013)\citenamefont
  {Bocquillon}, \citenamefont {Freulon}, \citenamefont {Berroir}, \citenamefont
  {Degiovanni}, \citenamefont {Pla{\c{c}}ais}, \citenamefont {Cavanna},
  \citenamefont {Jin},\ and\ \citenamefont {F{\`e}ve}}]{BocquillonSCS}%
  \BibitemOpen
  \bibfield  {author} {\bibinfo {author} {\bibfnamefont {E.}~\bibnamefont
  {Bocquillon}}, \bibinfo {author} {\bibfnamefont {V.}~\bibnamefont {Freulon}},
  \bibinfo {author} {\bibfnamefont {J.-.~M.}\ \bibnamefont {Berroir}}, \bibinfo
  {author} {\bibfnamefont {P.}~\bibnamefont {Degiovanni}}, \bibinfo {author}
  {\bibfnamefont {B.}~\bibnamefont {Pla{\c{c}}ais}}, \bibinfo {author}
  {\bibfnamefont {A.}~\bibnamefont {Cavanna}}, \bibinfo {author} {\bibfnamefont
  {Y.}~\bibnamefont {Jin}}, \ and\ \bibinfo {author} {\bibfnamefont
  {G.}~\bibnamefont {F{\`e}ve}},\ }\href {\doibase 10.1038/ncomms2788}
  {\bibfield  {journal} {\bibinfo  {journal} {Nature Communications}\ }\textbf
  {\bibinfo {volume} {4}},\ \bibinfo {pages} {1839} (\bibinfo {year}
  {2013})}\BibitemShut {NoStop}%
\bibitem [{\citenamefont {Hashisaka}\ \emph {et~al.}(2017)\citenamefont
  {Hashisaka}, \citenamefont {Hiyama}, \citenamefont {Akiho}, \citenamefont
  {Muraki},\ and\ \citenamefont {Fujisawa}}]{HashisakaSCS}%
  \BibitemOpen
  \bibfield  {author} {\bibinfo {author} {\bibfnamefont {M.}~\bibnamefont
  {Hashisaka}}, \bibinfo {author} {\bibfnamefont {N.}~\bibnamefont {Hiyama}},
  \bibinfo {author} {\bibfnamefont {T.}~\bibnamefont {Akiho}}, \bibinfo
  {author} {\bibfnamefont {K.}~\bibnamefont {Muraki}}, \ and\ \bibinfo {author}
  {\bibfnamefont {T.}~\bibnamefont {Fujisawa}},\ }\href {\doibase
  10.1038/nphys4062} {\bibfield  {journal} {\bibinfo  {journal} {Nature
  Physics}\ }\textbf {\bibinfo {volume} {13}},\ \bibinfo {pages} {559}
  (\bibinfo {year} {2017})}\BibitemShut {NoStop}%
\bibitem [{\citenamefont {Vekua}\ \emph {et~al.}(2009)\citenamefont {Vekua},
  \citenamefont {Matveenko},\ and\ \citenamefont {Shlyapnikov}}]{Vekua}%
  \BibitemOpen
  \bibfield  {author} {\bibinfo {author} {\bibfnamefont {T.}~\bibnamefont
  {Vekua}}, \bibinfo {author} {\bibfnamefont {S.~I.}\ \bibnamefont
  {Matveenko}}, \ and\ \bibinfo {author} {\bibfnamefont {G.~V.}\ \bibnamefont
  {Shlyapnikov}},\ }\href {\doibase 10.1134/S0021364009160139} {\bibfield
  {journal} {\bibinfo  {journal} {JETP Letters}\ }\textbf {\bibinfo {volume}
  {90}},\ \bibinfo {pages} {289} (\bibinfo {year} {2009})}\BibitemShut
  {NoStop}%
\bibitem [{\citenamefont {Brazovskii}\ \emph {et~al.}(1994)\citenamefont
  {Brazovskii}, \citenamefont {Matveenko},\ and\ \citenamefont
  {Nozieres}}]{Brazovskii}%
  \BibitemOpen
  \bibfield  {author} {\bibinfo {author} {\bibfnamefont {S.}~\bibnamefont
  {Brazovskii}}, \bibinfo {author} {\bibfnamefont {S.}~\bibnamefont
  {Matveenko}}, \ and\ \bibinfo {author} {\bibfnamefont {P.}~\bibnamefont
  {Nozieres}},\ }\href@noop {} {\bibfield  {journal} {\bibinfo  {journal}
  {Journal de Physique I}\ }\textbf {\bibinfo {volume} {4}},\ \bibinfo {pages}
  {571} (\bibinfo {year} {1994})}\BibitemShut {NoStop}%
\bibitem [{\citenamefont {Kimura}\ \emph {et~al.}(1996)\citenamefont {Kimura},
  \citenamefont {Kuroki},\ and\ \citenamefont {Aoki}}]{Aoki1996}%
  \BibitemOpen
  \bibfield  {author} {\bibinfo {author} {\bibfnamefont {T.}~\bibnamefont
  {Kimura}}, \bibinfo {author} {\bibfnamefont {K.}~\bibnamefont {Kuroki}}, \
  and\ \bibinfo {author} {\bibfnamefont {H.}~\bibnamefont {Aoki}},\ }\href
  {\doibase 10.1103/PhysRevB.53.9572} {\bibfield  {journal} {\bibinfo
  {journal} {Phys. Rev. B}\ }\textbf {\bibinfo {volume} {53}},\ \bibinfo
  {pages} {9572} (\bibinfo {year} {1996})}\BibitemShut {NoStop}%
\bibitem [{\citenamefont {Moroz}\ \emph
  {et~al.}(2000{\natexlab{a}})\citenamefont {Moroz}, \citenamefont {Samokhin},\
  and\ \citenamefont {Barnes}}]{MorozTheory}%
  \BibitemOpen
  \bibfield  {author} {\bibinfo {author} {\bibfnamefont {A.~V.}\ \bibnamefont
  {Moroz}}, \bibinfo {author} {\bibfnamefont {K.~V.}\ \bibnamefont {Samokhin}},
  \ and\ \bibinfo {author} {\bibfnamefont {C.~H.~W.}\ \bibnamefont {Barnes}},\
  }\href {\doibase 10.1103/PhysRevB.62.16900} {\bibfield  {journal} {\bibinfo
  {journal} {Phys. Rev. B}\ }\textbf {\bibinfo {volume} {62}},\ \bibinfo
  {pages} {16900} (\bibinfo {year} {2000}{\natexlab{a}})}\BibitemShut {NoStop}%
\bibitem [{\citenamefont {Moroz}\ \emph
  {et~al.}(2000{\natexlab{b}})\citenamefont {Moroz}, \citenamefont {Samokhin},\
  and\ \citenamefont {Barnes}}]{MorozExpr}%
  \BibitemOpen
  \bibfield  {author} {\bibinfo {author} {\bibfnamefont {A.~V.}\ \bibnamefont
  {Moroz}}, \bibinfo {author} {\bibfnamefont {K.~V.}\ \bibnamefont {Samokhin}},
  \ and\ \bibinfo {author} {\bibfnamefont {C.~H.~W.}\ \bibnamefont {Barnes}},\
  }\href {\doibase 10.1103/PhysRevLett.84.4164} {\bibfield  {journal} {\bibinfo
   {journal} {Phys. Rev. Lett.}\ }\textbf {\bibinfo {volume} {84}},\ \bibinfo
  {pages} {4164} (\bibinfo {year} {2000}{\natexlab{b}})}\BibitemShut {NoStop}%
\bibitem [{\citenamefont {Iucci}(2003)}]{Iucci}%
  \BibitemOpen
  \bibfield  {author} {\bibinfo {author} {\bibfnamefont {A.}~\bibnamefont
  {Iucci}},\ }\href {\doibase 10.1103/PhysRevB.68.075107} {\bibfield  {journal}
  {\bibinfo  {journal} {Phys. Rev. B}\ }\textbf {\bibinfo {volume} {68}},\
  \bibinfo {pages} {075107} (\bibinfo {year} {2003})}\BibitemShut {NoStop}%
\bibitem [{\citenamefont {Datta}\ and\ \citenamefont {Das}(1990)}]{Datta1990}%
  \BibitemOpen
  \bibfield  {author} {\bibinfo {author} {\bibfnamefont {S.}~\bibnamefont
  {Datta}}\ and\ \bibinfo {author} {\bibfnamefont {B.}~\bibnamefont {Das}},\
  }\href {\doibase 10.1063/1.102730} {\bibfield  {journal} {\bibinfo  {journal}
  {Applied Physics Letters}\ }\textbf {\bibinfo {volume} {56}},\ \bibinfo
  {pages} {665} (\bibinfo {year} {1990})},\ \Eprint
  {http://arxiv.org/abs/https://doi.org/10.1063/1.102730}
  {https://doi.org/10.1063/1.102730} \BibitemShut {NoStop}%
\bibitem [{\citenamefont {Nadj-Perge}\ \emph {et~al.}(2010)\citenamefont
  {Nadj-Perge}, \citenamefont {Frolov}, \citenamefont {Bakkers},\ and\
  \citenamefont {Kouwenhoven}}]{Frolov2010}%
  \BibitemOpen
  \bibfield  {author} {\bibinfo {author} {\bibfnamefont {S.}~\bibnamefont
  {Nadj-Perge}}, \bibinfo {author} {\bibfnamefont {S.~M.}\ \bibnamefont
  {Frolov}}, \bibinfo {author} {\bibfnamefont {E.~P. A.~M.}\ \bibnamefont
  {Bakkers}}, \ and\ \bibinfo {author} {\bibfnamefont {L.~P.}\ \bibnamefont
  {Kouwenhoven}},\ }\href {\doibase 10.1038/nature09682} {\bibfield  {journal}
  {\bibinfo  {journal} {Nature}\ }\textbf {\bibinfo {volume} {468}},\ \bibinfo
  {pages} {1084} (\bibinfo {year} {2010})}\BibitemShut {NoStop}%
\bibitem [{\citenamefont {Fasth}\ \emph {et~al.}(2007)\citenamefont {Fasth},
  \citenamefont {Fuhrer}, \citenamefont {Samuelson}, \citenamefont {Golovach},\
  and\ \citenamefont {Loss}}]{FasthSOC}%
  \BibitemOpen
  \bibfield  {author} {\bibinfo {author} {\bibfnamefont {C.}~\bibnamefont
  {Fasth}}, \bibinfo {author} {\bibfnamefont {A.}~\bibnamefont {Fuhrer}},
  \bibinfo {author} {\bibfnamefont {L.}~\bibnamefont {Samuelson}}, \bibinfo
  {author} {\bibfnamefont {V.~N.}\ \bibnamefont {Golovach}}, \ and\ \bibinfo
  {author} {\bibfnamefont {D.}~\bibnamefont {Loss}},\ }\href {\doibase
  10.1103/PhysRevLett.98.266801} {\bibfield  {journal} {\bibinfo  {journal}
  {Phys. Rev. Lett.}\ }\textbf {\bibinfo {volume} {98}},\ \bibinfo {pages}
  {266801} (\bibinfo {year} {2007})}\BibitemShut {NoStop}%
\bibitem [{\citenamefont {Sato}\ \emph {et~al.}(2019)\citenamefont {Sato},
  \citenamefont {Matsuo}, \citenamefont {Hsu}, \citenamefont {Stano},
  \citenamefont {Ueda}, \citenamefont {Takeshige}, \citenamefont {Kamata},
  \citenamefont {Lee}, \citenamefont {Shojaei}, \citenamefont {Wickramasinghe},
  \citenamefont {Shabani}, \citenamefont {Palmstr\o{}m}, \citenamefont
  {Tokura}, \citenamefont {Loss},\ and\ \citenamefont
  {Tarucha}}]{SatoExperiment}%
  \BibitemOpen
  \bibfield  {author} {\bibinfo {author} {\bibfnamefont {Y.}~\bibnamefont
  {Sato}}, \bibinfo {author} {\bibfnamefont {S.}~\bibnamefont {Matsuo}},
  \bibinfo {author} {\bibfnamefont {C.-H.}\ \bibnamefont {Hsu}}, \bibinfo
  {author} {\bibfnamefont {P.}~\bibnamefont {Stano}}, \bibinfo {author}
  {\bibfnamefont {K.}~\bibnamefont {Ueda}}, \bibinfo {author} {\bibfnamefont
  {Y.}~\bibnamefont {Takeshige}}, \bibinfo {author} {\bibfnamefont
  {H.}~\bibnamefont {Kamata}}, \bibinfo {author} {\bibfnamefont {J.~S.}\
  \bibnamefont {Lee}}, \bibinfo {author} {\bibfnamefont {B.}~\bibnamefont
  {Shojaei}}, \bibinfo {author} {\bibfnamefont {K.}~\bibnamefont
  {Wickramasinghe}}, \bibinfo {author} {\bibfnamefont {J.}~\bibnamefont
  {Shabani}}, \bibinfo {author} {\bibfnamefont {C.}~\bibnamefont
  {Palmstr\o{}m}}, \bibinfo {author} {\bibfnamefont {Y.}~\bibnamefont
  {Tokura}}, \bibinfo {author} {\bibfnamefont {D.}~\bibnamefont {Loss}}, \ and\
  \bibinfo {author} {\bibfnamefont {S.}~\bibnamefont {Tarucha}},\ }\href
  {\doibase 10.1103/PhysRevB.99.155304} {\bibfield  {journal} {\bibinfo
  {journal} {Phys. Rev. B}\ }\textbf {\bibinfo {volume} {99}},\ \bibinfo
  {pages} {155304} (\bibinfo {year} {2019})}\BibitemShut {NoStop}%
\bibitem [{\citenamefont {Hsu}\ \emph {et~al.}(2019)\citenamefont {Hsu},
  \citenamefont {Stano}, \citenamefont {Sato}, \citenamefont {Matsuo},
  \citenamefont {Tarucha},\ and\ \citenamefont {Loss}}]{SatoTheory}%
  \BibitemOpen
  \bibfield  {author} {\bibinfo {author} {\bibfnamefont {C.-H.}\ \bibnamefont
  {Hsu}}, \bibinfo {author} {\bibfnamefont {P.}~\bibnamefont {Stano}}, \bibinfo
  {author} {\bibfnamefont {Y.}~\bibnamefont {Sato}}, \bibinfo {author}
  {\bibfnamefont {S.}~\bibnamefont {Matsuo}}, \bibinfo {author} {\bibfnamefont
  {S.}~\bibnamefont {Tarucha}}, \ and\ \bibinfo {author} {\bibfnamefont
  {D.}~\bibnamefont {Loss}},\ }\href {\doibase 10.1103/PhysRevB.100.195423}
  {\bibfield  {journal} {\bibinfo  {journal} {Phys. Rev. B}\ }\textbf {\bibinfo
  {volume} {100}},\ \bibinfo {pages} {195423} (\bibinfo {year}
  {2019})}\BibitemShut {NoStop}%
\bibitem [{\citenamefont {Hamamoto}\ \emph {et~al.}(2008)\citenamefont
  {Hamamoto}, \citenamefont {Imura},\ and\ \citenamefont
  {Kato}}]{NumericalStudyPhasePortrait}%
  \BibitemOpen
  \bibfield  {author} {\bibinfo {author} {\bibfnamefont {Y.}~\bibnamefont
  {Hamamoto}}, \bibinfo {author} {\bibfnamefont {K.-I.}\ \bibnamefont {Imura}},
  \ and\ \bibinfo {author} {\bibfnamefont {T.}~\bibnamefont {Kato}},\ }\href
  {\doibase 10.1103/PhysRevB.77.165402} {\bibfield  {journal} {\bibinfo
  {journal} {Phys. Rev. B}\ }\textbf {\bibinfo {volume} {77}},\ \bibinfo
  {pages} {165402} (\bibinfo {year} {2008})}\BibitemShut {NoStop}%
\bibitem [{\citenamefont {Hikihara}\ \emph {et~al.}(2005)\citenamefont
  {Hikihara}, \citenamefont {Furusaki},\ and\ \citenamefont
  {Matveev}}]{Matveev}%
  \BibitemOpen
  \bibfield  {author} {\bibinfo {author} {\bibfnamefont {T.}~\bibnamefont
  {Hikihara}}, \bibinfo {author} {\bibfnamefont {A.}~\bibnamefont {Furusaki}},
  \ and\ \bibinfo {author} {\bibfnamefont {K.~A.}\ \bibnamefont {Matveev}},\
  }\href {\doibase 10.1103/PhysRevB.72.035301} {\bibfield  {journal} {\bibinfo
  {journal} {Phys. Rev. B}\ }\textbf {\bibinfo {volume} {72}},\ \bibinfo
  {pages} {035301} (\bibinfo {year} {2005})}\BibitemShut {NoStop}%
\bibitem [{\citenamefont {Kamide}\ \emph {et~al.}(2006)\citenamefont {Kamide},
  \citenamefont {Tsukada},\ and\ \citenamefont
  {Kurihara}}]{KimuraResonantZeeman}%
  \BibitemOpen
  \bibfield  {author} {\bibinfo {author} {\bibfnamefont {K.}~\bibnamefont
  {Kamide}}, \bibinfo {author} {\bibfnamefont {Y.}~\bibnamefont {Tsukada}}, \
  and\ \bibinfo {author} {\bibfnamefont {S.}~\bibnamefont {Kurihara}},\ }\href
  {\doibase 10.1103/PhysRevB.73.235326} {\bibfield  {journal} {\bibinfo
  {journal} {Phys. Rev. B}\ }\textbf {\bibinfo {volume} {73}},\ \bibinfo
  {pages} {235326} (\bibinfo {year} {2006})}\BibitemShut {NoStop}%
\bibitem [{\citenamefont {Governale}\ and\ \citenamefont
  {Zülicke}(2004)}]{Governale}%
  \BibitemOpen
  \bibfield  {author} {\bibinfo {author} {\bibfnamefont {M.}~\bibnamefont
  {Governale}}\ and\ \bibinfo {author} {\bibfnamefont {U.}~\bibnamefont
  {Zülicke}},\ }\href {\doibase https://doi.org/10.1016/j.ssc.2004.05.047}
  {\bibfield  {journal} {\bibinfo  {journal} {Solid State Communications}\
  }\textbf {\bibinfo {volume} {131}},\ \bibinfo {pages} {581} (\bibinfo {year}
  {2004})},\ \bibinfo {note} {new advances on collective phenomena in
  one-dimensional systems}\BibitemShut {NoStop}%
\bibitem [{\citenamefont {Governale}\ and\ \citenamefont
  {Z\"ulicke}(2002)}]{Zulicke}%
  \BibitemOpen
  \bibfield  {author} {\bibinfo {author} {\bibfnamefont {M.}~\bibnamefont
  {Governale}}\ and\ \bibinfo {author} {\bibfnamefont {U.}~\bibnamefont
  {Z\"ulicke}},\ }\href {\doibase 10.1103/PhysRevB.66.073311} {\bibfield
  {journal} {\bibinfo  {journal} {Phys. Rev. B}\ }\textbf {\bibinfo {volume}
  {66}},\ \bibinfo {pages} {073311} (\bibinfo {year} {2002})}\BibitemShut
  {NoStop}%
\bibitem [{\citenamefont {Gritsev}\ \emph {et~al.}(2005)\citenamefont
  {Gritsev}, \citenamefont {Japaridze}, \citenamefont {Pletyukhov},\ and\
  \citenamefont {Baeriswyl}}]{Gritsev}%
  \BibitemOpen
  \bibfield  {author} {\bibinfo {author} {\bibfnamefont {V.}~\bibnamefont
  {Gritsev}}, \bibinfo {author} {\bibfnamefont {G.}~\bibnamefont {Japaridze}},
  \bibinfo {author} {\bibfnamefont {M.}~\bibnamefont {Pletyukhov}}, \ and\
  \bibinfo {author} {\bibfnamefont {D.}~\bibnamefont {Baeriswyl}},\ }\href
  {\doibase 10.1103/PhysRevLett.94.137207} {\bibfield  {journal} {\bibinfo
  {journal} {Phys. Rev. Lett.}\ }\textbf {\bibinfo {volume} {94}},\ \bibinfo
  {pages} {137207} (\bibinfo {year} {2005})}\BibitemShut {NoStop}%
\bibitem [{\citenamefont {Schulz}\ \emph {et~al.}(2009)\citenamefont {Schulz},
  \citenamefont {De~Martino}, \citenamefont {Ingenhoven},\ and\ \citenamefont
  {Egger}}]{Schulz}%
  \BibitemOpen
  \bibfield  {author} {\bibinfo {author} {\bibfnamefont {A.}~\bibnamefont
  {Schulz}}, \bibinfo {author} {\bibfnamefont {A.}~\bibnamefont {De~Martino}},
  \bibinfo {author} {\bibfnamefont {P.}~\bibnamefont {Ingenhoven}}, \ and\
  \bibinfo {author} {\bibfnamefont {R.}~\bibnamefont {Egger}},\ }\href
  {\doibase 10.1103/PhysRevB.79.205432} {\bibfield  {journal} {\bibinfo
  {journal} {Phys. Rev. B}\ }\textbf {\bibinfo {volume} {79}},\ \bibinfo
  {pages} {205432} (\bibinfo {year} {2009})}\BibitemShut {NoStop}%
\bibitem [{\citenamefont {Kainaris}\ and\ \citenamefont
  {Carr}(2015)}]{Kainaris}%
  \BibitemOpen
  \bibfield  {author} {\bibinfo {author} {\bibfnamefont {N.}~\bibnamefont
  {Kainaris}}\ and\ \bibinfo {author} {\bibfnamefont {S.~T.}\ \bibnamefont
  {Carr}},\ }\href {\doibase 10.1103/PhysRevB.92.035139} {\bibfield  {journal}
  {\bibinfo  {journal} {Phys. Rev. B}\ }\textbf {\bibinfo {volume} {92}},\
  \bibinfo {pages} {035139} (\bibinfo {year} {2015})}\BibitemShut {NoStop}%
\bibitem [{\citenamefont {Kainaris}\ \emph {et~al.}(2018)\citenamefont
  {Kainaris}, \citenamefont {Carr},\ and\ \citenamefont
  {Mirlin}}]{KainarisImp}%
  \BibitemOpen
  \bibfield  {author} {\bibinfo {author} {\bibfnamefont {N.}~\bibnamefont
  {Kainaris}}, \bibinfo {author} {\bibfnamefont {S.~T.}\ \bibnamefont {Carr}},
  \ and\ \bibinfo {author} {\bibfnamefont {A.~D.}\ \bibnamefont {Mirlin}},\
  }\href {\doibase 10.1103/PhysRevB.97.115107} {\bibfield  {journal} {\bibinfo
  {journal} {Phys. Rev. B}\ }\textbf {\bibinfo {volume} {97}},\ \bibinfo
  {pages} {115107} (\bibinfo {year} {2018})}\BibitemShut {NoStop}%
\bibitem [{\citenamefont {Fisher}\ and\ \citenamefont
  {Zwerger}(1985)}]{Zwerger1985}%
  \BibitemOpen
  \bibfield  {author} {\bibinfo {author} {\bibfnamefont {M.~P.~A.}\
  \bibnamefont {Fisher}}\ and\ \bibinfo {author} {\bibfnamefont
  {W.}~\bibnamefont {Zwerger}},\ }\href {\doibase 10.1103/PhysRevB.32.6190}
  {\bibfield  {journal} {\bibinfo  {journal} {Phys. Rev. B}\ }\textbf {\bibinfo
  {volume} {32}},\ \bibinfo {pages} {6190} (\bibinfo {year}
  {1985})}\BibitemShut {NoStop}%
\bibitem [{\citenamefont {Caldeira}\ and\ \citenamefont
  {Leggett}(1983)}]{Caldeira1983}%
  \BibitemOpen
  \bibfield  {author} {\bibinfo {author} {\bibfnamefont {A.}~\bibnamefont
  {Caldeira}}\ and\ \bibinfo {author} {\bibfnamefont {A.}~\bibnamefont
  {Leggett}},\ }\href {\doibase https://doi.org/10.1016/0003-4916(83)90202-6}
  {\bibfield  {journal} {\bibinfo  {journal} {Annals of Physics}\ }\textbf
  {\bibinfo {volume} {149}},\ \bibinfo {pages} {374} (\bibinfo {year}
  {1983})}\BibitemShut {NoStop}%
\end{thebibliography}%
   \end{document}